\documentclass[twocolumn]{aastex63}
\usepackage{natbib}
\usepackage{color}
\usepackage{mathtools}
\usepackage{hyperref} 
\hypersetup{colorlinks=true,citecolor=blue}

\def    \apjl  		{\rm {ApJL}}

\def    \apjl  		{\rm {ApJL}}

\def	\cm		{\,{\rm {cm}}}
\def	\K		{\,{\rm K}}
\def	\g		{\,{\rm {g}}}
\def	\mum	{\,{\mu \rm{m}}}

\def \bea {\begin{eqnarray}}
\def \ena {\end{eqnarray}}

\def	\B	{{\rm B}}

\def	\C	{{\rm C}}

\def	\cm	{\,{\rm cm}}
\def	\d	{{\rm d}}

\def	\D	{{\rm D}}

\def	\erg	{\,{\rm erg}}

\def	\g	{\,{\rm g}}
\def	\gas	{\,{\rm gas}}
\def	\km	{\,{\rm km}}
\def	\G	{{\rm G}}

\def	\H	{{\rm H}}

\def    \kB    {k_{\rm B}}

\def	\N	{{\rm N}}

\def	\s	{\,{\rm s}}

\def	\AU	{\,{\rm AU}}

\def	\V	{{\rm V}}

\def	\rad	{\,{\rm rad}}
\def	\yr	    {\,{\rm yr}}

\def    \gas     	{{\rm gas}}

\begin{document}
\shorttitle{Rotational disruption of F-corona dust}
\shortauthors{Hoang et al.}
\title{Effect of Dust Rotational Disruption by Radiative Torques and Implications for F-corona decrease revealed by the Parker Solar Probe}

\author{Thiem Hoang}
\affil{Korea Astronomy and Space Science Institute, Daejeon 34055, Republic of Korea}
\affil{University of Science and Technology, Korea, (UST), 217 Gajeong-ro Yuseong-gu, Daejeon 34113, Republic of Korea}

\author{Alex Lazarian}
\affil{Department of Astronomy, University of Wisconsin-Madison, USA}
\affil{Korea Astronomy and Space Science Institute, Daejeon 34055, Republic of Korea}

\author{Hyeseung Lee}
\affil{Korea Astronomy and Space Science Institute, Daejeon 34055, Republic of Korea}

\author{Kyungsuk Cho}
\affil{Korea Astronomy and Space Science Institute, Daejeon 34055, Republic of Korea}
\affil{University of Science and Technology, Korea, (UST), 217 Gajeong-ro Yuseong-gu, Daejeon 34113, Republic of Korea}

\author{Pin-Gao Gu}
\affil{Institute of Astronomy and Astrophysics, Academia Sinica, 10617 Taipei, Taiwan}

\author{Chi-Hang Ng}
\affil{Institute of Astronomy and Astrophysics, Academia Sinica, 10617 Taipei, Taiwan}
\affil{Department of Physics, National Taiwan University, 10617 Taipei, Taiwan}

\begin{abstract}
The first-year results from the Parker Solar Probe (PSP) reveal a gradual decrease of F-coronal dust from distances of $D=0.166-0.336$ AU (or the inner elongations of $\sim 9.22- 18.69~R_{\odot}$) to the Sun \citep{Howard:2019ih}. Such a F-corona decrease cannot be explained by the dust sublimation scenario of the popular silicate composition that implies a dust-free-zone of boundary at heliocentric radius $R\lesssim 4-5R_{\odot}$, but may be explained by appealing to various dust compositions with different sublimation fronts. In this paper, we present an additional explanation for the F-corona decrease using our newly introduced mechanism of dust destruction so-called Radiative Torque Disruption (RATD) mechanism. We demonstrate that RATD rapidly breaks large grains into nanoparticles so that they can be efficiently destroyed by nonthermal sputtering induced by bombardment of energetic protons from slow solar winds, which extends the dust-free-zone established by thermal sublimation to $R_{dfz}\sim 8R_{\odot}$. Beyond this extended dust-free-zone, we find that the dust mass decreases gradually from $R\sim 42R_{\odot}$ toward the Sun due to partial removal of nanodust by nonthermal sputtering. The joint effect of RATD and nonthermal sputtering can successfully reproduce the gradual decrease of the F-corona between $19-9R_{\odot}$ observed by the PSP. Finally, the RATD mechanism can efficiently produce nanoparticles usually observed in the inner solar system. 
 
\end{abstract}

\section{Introduction}\label{sec:intro}
Dust grains orbiting the Sun gradually spiral inward due to the loss of their angular momentum via the Poynting-Robertson (P-R) drag. When dust grains are sufficiently close to the Sun, they are sublimated because the grain temperature can exceed the sublimation threshold, producing a dust-free zone around the Sun \citep{1929ApJ....69...49R}. 

\cite{1979PASJ...31..585M} quantified the radius of dust-free-zone by studying dust sublimation of various dust compositions and found that dust grains are completely evaporated within a heliocentric distance of $4R_{\odot}$ (see also \citealt{1998EP&S...50..493K} for a review). An exact location of dust-free-zone depends on grain compositions and grain sizes (see, e.g., \citealt{Kobayashi:2012fi}). 

Significant efforts have been made to observe the dust-free-zone from the ground during the solar eclipse (\citealt{1992Sci...257.1377L}) and from space (e.g., \citealt{1978A&A....64..119L}). Near-infrared observations by \cite{1992Sci...257.1377L} found no evidence of the predicted dust-free-zone, but other near-infrared observations reported the clear detection of the circumsolar dust ring at the outer edge of the dust-free zones (see \citealt{1998EP&S...50..493K} and references therein). First-year results from first two perihelion passes by the Parker Solar Probe (PSP) spacecraft, at heliocentric distances of $D=0.166-0.336$ AU from the Sun, reveal the decrease in the intensity of scattered light by circumsolar dust, so-called F-corona at small elongations ($\epsilon\sim 15-20^{\circ}$) \citep{Howard:2019ih}. The corresponding elongation in solar radii is $R_{\epsilon}\sim (0.166-0.336)\sin(15^{\circ})\AU\sim 9-19R_{\odot}$, which is suggested to see the outer edge of dust free zone (Kimura private communication). Extrapolation of the thinning trend of the F-corona suggests the possibility of observing the true dust-free-zone in the sixth flyby. In the next few years, the PSP will come closer to the Sun and reach as close as $D\sim 10R_{\odot}$ ($R_{\epsilon}\sim 2.6R_{\odot}$), which will shed light on the dust-free-zone (\citealt{2016SSRv..204....7F}). 

The gradual thinning out of the F-corona observed by the PSP between elongations of $R_{\epsilon}\sim 19-9 R_{\odot}$ is difficult to explain with the sublimation model of the standard interstellar dust that implies a sharp radius at $\sim 4-5R_{\odot}$ (see \citealt{1992A&A...261..329M}). However, it could be explained by appealing to the dependence of the sublimation zone on the composition of dust particles (e.g., see \citealt{2004SSRv..110..269M}). For instance, magnesium-rich pyroxene grains are expected to sublime around $5-6R_{\odot}$, while magnesium-rich olivine grains suffer thermal sublimation already at $\sim 10-16 R_{\odot}$ (see \citealt{Kimura:2002bb}). Iron-rich olivine grains most likely sublime at even larger heliocentric distances, owing to their high absorptivity in the visible wavelength range. Sublimation of metallic grains is expected to start at $\sim 25R_{\odot}$ (\citealt{1974A&A....35..197L}).

In this paper, we aim to present an additional explanation for the F-corona decrease by considering the new effect of RAdiative Torque Disruption (RATD), which is discovered by \cite{Hoang:2019da} (see also \citealt{2019ApJ...876...13H}). The RATD mechanism is based on the fact that dust grains of irregular shapes exposed to anisotropic radiation field experience Radiative Torques (RATs; \citealt{Dolginov:1976p2480}; \citealt{1996ApJ...470..551D}; \citealt{2007MNRAS.378..910L}; \citealt{Hoang:2008gb}). RATs can spin up the grain to suprathermal rotation (\citealt{1996ApJ...470..551D}; \citealt{2004ApJ...614..781A}) such that the resulting centrifugal stress can exceed the maximum tensile strength of the grain material, which breaks the grain into small fragments (\citealt{Hoang:2019da}). 
We expect that RATD rapidly breaks micron-sized grains into nanoparticles, such that the resulting nanoparticles can be destroyed rapidly by nonthermal sputtering (e.g., \citealt{1994ApJ...433..797J}) due to energetic protons of the solar wind. Indeed, the effect of RATD on enhanced thermal sublimation of icy grains was first explored in \cite{Hoang:2020hn} in star-forming regions. The similar effect is expected for dust grains in the circumsolar region. Note that, previously, \cite{1969JGR....74.4379P} suggested rotational bursting of interplanetary dust by radiative torques, but his treatment is based on a rough estimate using radiation pressure and did not study the dependence on the radiation wavelength, grain shape, grain size, and composition. Our study here is based on RATs which are numerically calculated from a large sample of grain shapes and compositions and radiation wavelengths using DDSCAT (\citealt{1994JOSAA..11.1491D}; \citealt{1996ApJ...470..551D}) and T-matrix codes (\citealt{{2019ApJ...878...96H},{Herranen.2021}}). We also take consider rotational damping by gas and infrared emission, which is ignored in \cite{1969JGR....74.4379P}.

The structure of this paper is as follows. In Section \ref{sec:theory}, we study the disruption of dust grains into nanoparticles by RATD induced by solar radiation and calculate the maximum grain size as a function of the heliocentric distance. In Section \ref{sec:nano}, we study the destruction of nanoparticles by nonthermal sputtering induced by the solar wind and calculate the mass loss of dust due to sputtering. We discuss our results for observations and in-situ measurements by the PSP in Section \ref{sec:discuss}. A short summary of our main findings is presented in Section \ref{sec:sum}.

\section{Rotational disruption of dust grains by Radiative Torques}
\label{sec:theory}
In this section, we will use the RATD mechanism to determine the upper cutoff of the grain size distribution in the F-corona.

\subsection{Basic Theory of Rotational Disruption}

A dust grain of radius $a$ rotating at velocity $\omega$ develops a centrifugal stress due to centrifugal force, which scales as $S=\rho a^{2} \omega^{2}/4$ with $\rho$ being the mass density of grain material (\citealt{Hoang:2019da}). When the rotation rate increases to a critical limit such that the tensile stress induced by centrifugal force exceeds the maximum tensile stress, the so-called tensile strength of the material ($S_{\max}$), the grain is disrupted instantaneously. The critical angular velocity for the disruption is given by
\bea
\omega_{\rm cri}&=&\frac{2}{a}\left(\frac{S_{\rm max}}{\rho} \right)^{1/2}\nonumber\\
&\simeq& 3.6\times 10^{9}a_{-5}^{-1}\hat{\rho}^{-1/2}S_{\rm max,9}^{1/2}~\rm rad/s,~~~~\label{eq:omega_cri}
\ena
where $a_{-5}=a/(10^{-5}\cm)$, $\hat{\rho}=\rho/(3\g\cm^{-3})$, $S_{\rm max}$ is the tensile strength of dust material and $S_{\rm max,9}=S_{\rm max}/(10^{9}\erg\cm^{-3})$ is the tensile strength in units of $10^{9}\erg\cm^{-3}$.

The exact value of $S_{\max}$ depends on the dust grain composition and structure. Compact grains have higher $S_{\max}$ than porous/composite grains. Ideal material without impurity, such as diamond, can have $S_{\max}\ge 10^{11}\erg\cm^{-3}$ (\citealt{1974ApJ...190....1B}; see \citealt{Hoang:2019da} for more details), while dust aggregates have lower tensile strength (see \citealt{2019ApJ...874..159T}; \citealt{2020MNRAS.496.1667K} for more details). In the following, we consider a range of the tensile strength from $S_{\max} = 10^{8}-10^{11}\erg\cm^{-3}$. The largest $S_{\max}$ is expected for nanoparticles that are likely to have compact structures.

\subsection{Grain Rotation Rate induced by Solar radiation}\label{sec:results}
Let $u_{\lambda}$ be the spectral energy density of radiation field at wavelength $\lambda$. To describe the strength of a radiation field, let define $U=u_{\rm rad}/u_{\rm ISRF}$ with $u_{\rm ISRF}=8.64\times 10^{-13}\erg\cm^{-3}$ being the energy density of the average interstellar radiation field (ISRF) in the solar neighborhood as given by \cite{1983A&A...128..212M}. 

The radiation energy density at distance $R$ from the Sun is given by
\bea
u_{\rm rad}&=&\frac{L_{\star}}{4\pi R^{2}c}\\ \nonumber
&\simeq & 4.5\times 10^{-5}\left(\frac{L_{\star}}{L_{\odot}}\right)\left(\frac{R}{1\AU}\right)^{-2}\erg\cm^{-3},
\ena
where $L_{\star}=L_{\odot}$ for the solar bolometric luminosity.

The radiation strength at distance $R$ becomes
\bea
U(R)\simeq U_{0}\left(\frac{L_{\star}}{L_{\odot}}\right)\left(\frac{R}{1\AU}\right)^{-2},
\ena
where $U_{0}=5.2\times 10^{7}$ is the radiation strength at $R_{0}=1\AU$.

The mean wavelength of the solar radiation field is defined as
\bea
\bar{\lambda}=\frac{\int u_{\lambda}\lambda d\lambda}{\int u_{\lambda}d\lambda}
\ena
which yields $\bar{\lambda}\approx 0.91\mum$ for the solar-type star.

Grains subject to the anisotropic radiation field experience RATs which act to spin-up the grains to suprathermal rotation (\citealt{Dolginov:1976p2480}; \citealt{1996ApJ...470..551D}; \citealt{2007MNRAS.378..910L}; \citealt{Hoang:2008gb}). At the same time, grain rotation experiences damping due to collisions with gas species (of proton density $n_{\H}$ and temperature $T_{\rm gas}$) and infrared emission (see Appendix \ref{sec:RATD}). The grain rotation velocity induced by RATs is given by (\citealt{Hoang:2019da} and \citealt{2019ApJ...876...13H})
\bea
\omega_{\rm RAT}&\simeq &3.2\times 10^{8}\gamma a_{-5}^{0.7}\bar{\lambda}_{0.5}^{-1.7}\nonumber\\
&\times&\left(\frac{U}{\hat{n}\hat{T}_{\rm gas}^{1/2}}\right)\left(\frac{1}{1+F_{\rm IR}}\right)\rad\s^{-1},~~~\label{eq:omega_RAT}
\ena
for grains with $a\lesssim a_{\rm trans}$, and
\bea
\omega_{\rm RAT}&\simeq &1.6\times 10^{9}\frac{\gamma}{a_{-5}^{2}}\bar{\lambda}_{0.5}\nonumber\\
&&\times \left(\frac{U}{\hat{n}\hat{T}_{\rm gas}^{1/2}}\right)\left(\frac{1}{1+F_{\rm IR}}\right)\rad\s^{-1},~~~
\ena
for grains with $a> a_{\rm trans}$ where $a_{\rm trans}\sim \bar{\lambda}/1.8$ is the transition grain size where RAT efficiency $Q_{\Gamma}$ changes the slope from $\eta\sim -3$ to $\eta=0$ (\citealt{2007MNRAS.378..910L}; see Appendix \ref{sec:RATD}). 

In the above equations, $\overline{\lambda}_{0.5}=\overline{\lambda}/0.5\mum$, $\hat{n}=n_{H}/10\cm^{-3}, \hat{T}_{\rm gas}=T_{\rm gas}/100\K$, $F_{\rm IR}$ is the dimensionless parameter that describes the grain rotational damping by infrared emission. In addition to gas and IR damping, grains experience rotational damping due to plasma drag and ion collisions (\citealt{1998ApJ...508..157D}; \citealt{Hoang:2010jy}). However, subject to intense solar radiation field, the infrared damping is dominant ($F_{\rm IR}\gg 1$), such that the grain rotation rate does not depend on the local gas properties (see Appendix \ref{sec:RATD} for more details).

\subsection{Grain disruption size and time vs. heliocentric distance}

For strong solar radiation fields considered in this paper, $F_{\rm IR}\gg 1$, from Equations (\ref{eq:omega_RAT}) and (\ref{eq:omega_cri}), one can obtain the disruption grain size:
\bea
a_{\rm disr}&\simeq&0.18\left(\gamma^{-1}U^{-1/3}\bar{\lambda}_
{0.5}^{1.7}S_{\max,9}^{1/2}\right)^{1/2.7}\mum,~~~~~\label{eq:adisr_comp1}
\ena
for $a_{\rm disr}\le a_{\rm trans}$. 

At $R=1\AU$, one obtains $a_{\rm disr}\sim 0.02\mum$ for the typical parameters of $S_{\max,9}=1$, $\gamma=1$, and $\lambda=0.91\mum$. At $R=0.1\AU$ ($21R_{\odot}$, one has $a_{\rm disr}=0.011\mum$. At $R=0.05\AU$  ($10R_{\odot}$), $a_{\rm disr}\sim 0.0096\mum$. At $R=0.02\AU~ (4R_{\odot}$), $a_{\rm disr}=0.0076\mum$. So, within $R=0.5\AU$, nanoparticles of size $a<a_{\rm disr}/2\sim 0.0055\mum$ (5.5 nm) are abundant due to RATD.

Due to the decrease of $\omega_{\rm RAT}$ with the grain size for $a>a_{\rm trans}$ (see Eq. \ref{eq:omega_RAT}), very large grains would not be disrupted. The maximum size of grains that can still be disrupted by RATD is given by (\citealt{2019ApJ...876...13H})
\bea
a_{\rm disr,max}&\simeq& 9.6\gamma\bar{\lambda}_{0.5}\left(\frac{U}{\hat{n}\hat{T}_{\gas}^{1/2}}\right)^{1/2}\left(\frac{1}{1+F_{\rm IR}}\right)\nonumber\\
&&\times \rho S_{\max,9}^{-1/2}~\mum.\label{eq:adisr_up}
\ena 

\begin{figure*}
\includegraphics[width=0.5\textwidth]{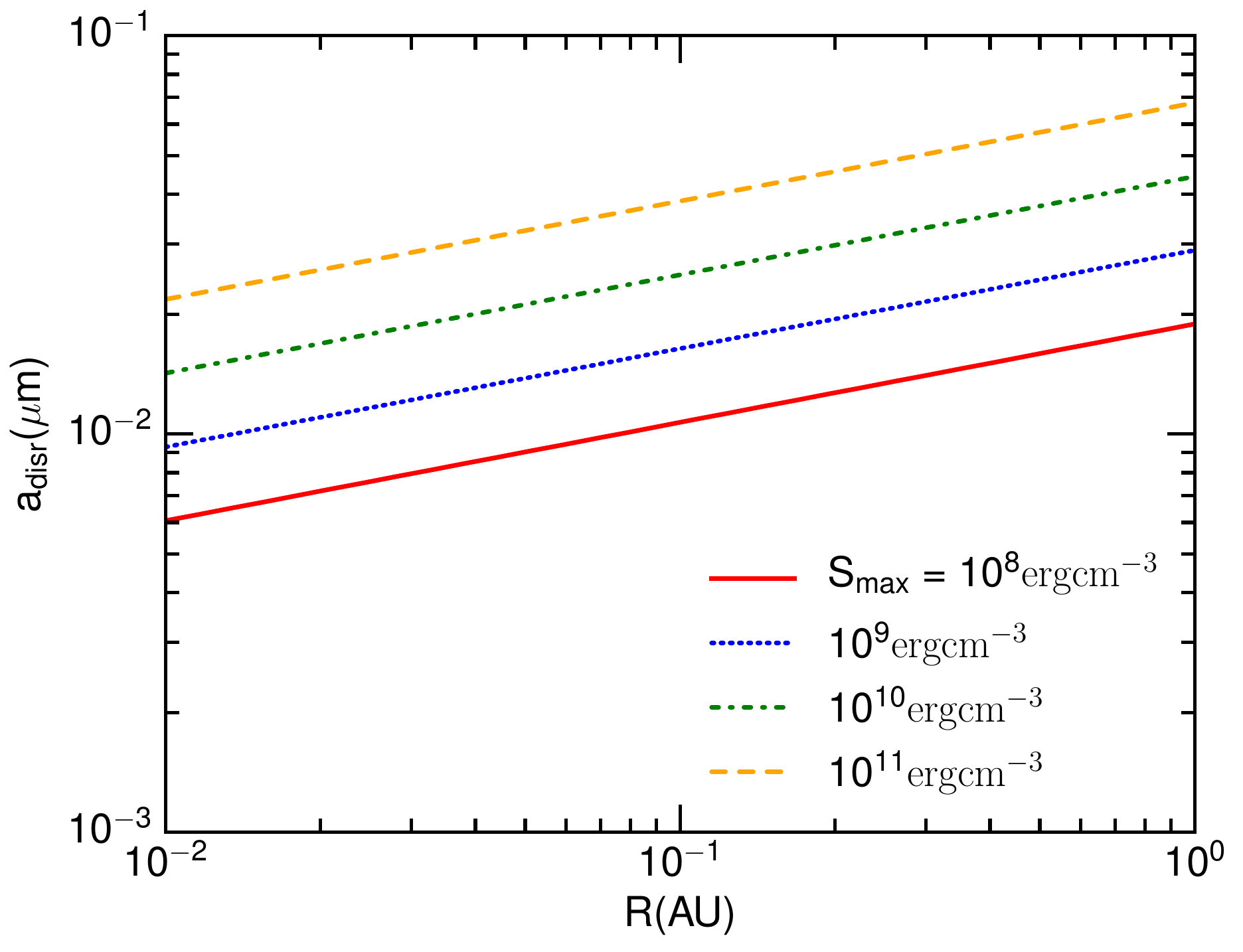}
\includegraphics[width=0.5\textwidth]{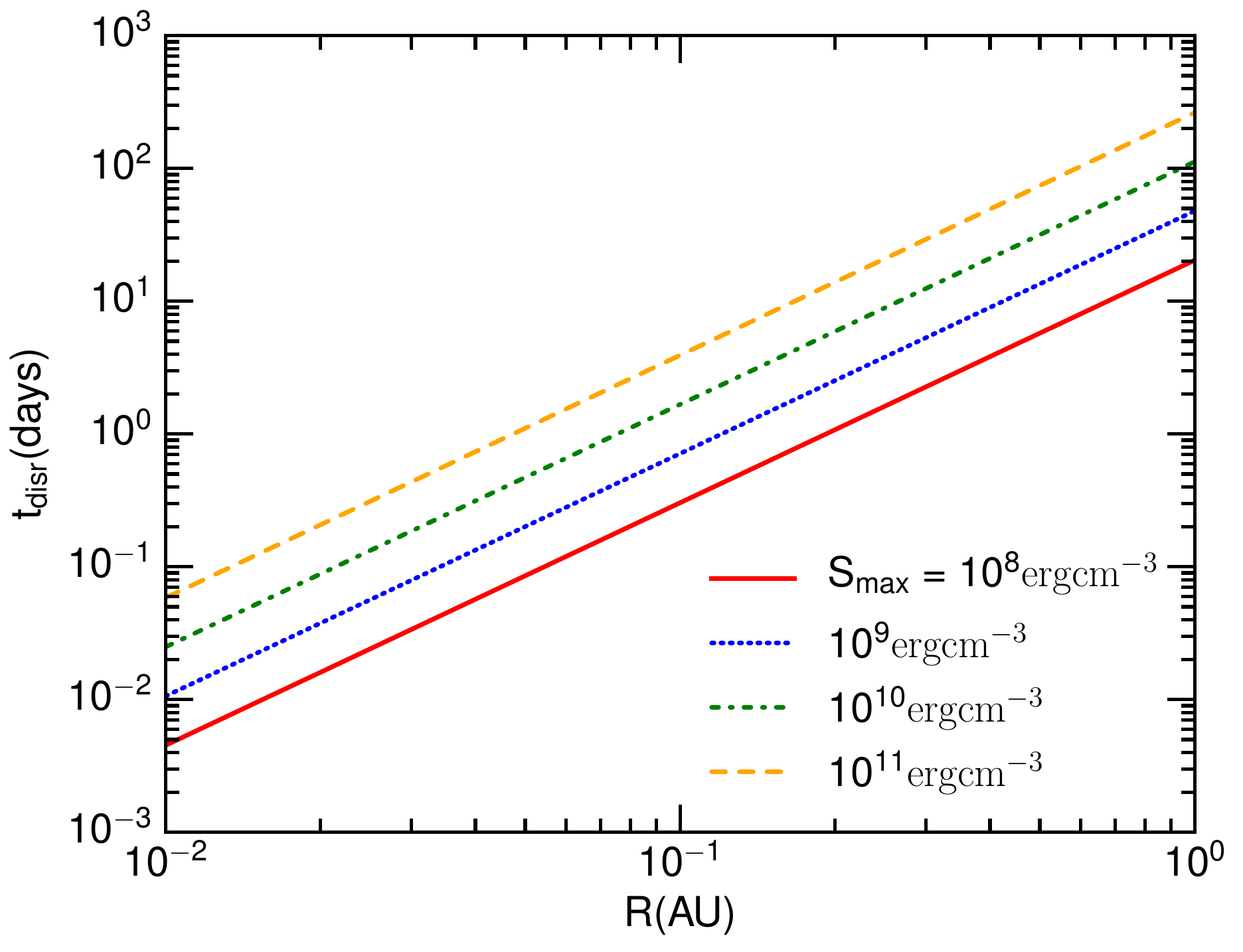}
\caption{Left panel: Grain disruption size, $a_{\rm disr}$, as a function of the heliocentric distance for the different tensile strengths. Disruption size is smaller for weaker grains (i.e., lower $S_{\rm max}$). Right panel: disruption time evaluated at $a=a_{\rm disr}$. Large grains of $a>0.01\mum$ are disrupted rapidly on a timescale of $t_{\rm disr}< 10$ days.}
\label{fig:adisr}
\end{figure*}

Figure \ref{fig:adisr} (upper panel) shows the grain disruption size vs. the heliocentric distance for the different tensile strength. For $R<5R_{\odot}$ (0.24 AU), only very small grains of sizes $a<0.005\mum$ (5 nm) can survive against RATD, assuming compact grains of $S_{\rm max}\sim 10^{9}\erg\cm^{-3}$. Ideal nanoparticles with $S_{\rm max}\sim 10^{11}\erg\cm^{-3}$ can have a maximum size of $a<0.03\mum$ (30 nm). For $R<50R_{\odot}$ (0.5 AU), we predict only small grains of $a<0.025\mum$ (25 nm). 

The characteristic timescale for rotational desorption can be estimated as (\citealt{Hoang:2019da}):
\bea
t_{\rm disr,0}&=&\frac{I\omega_{\rm disr}}{dJ/dt}=\frac{I\omega_{\rm disr}}{\Gamma_{\rm RAT}}\nonumber\\
&\simeq& 36.5 (\gamma U_{7})^{-1}\bar{\lambda}_{0.5}^{1.7}\hat{\rho}^{1/2}S_{\max,9}^{1/2}a_{-5}^{-0.7}{~\rm days}\label{eq:tdisr}
\ena
for $a_{\rm disr}<a \lesssim a_{\rm trans}$, and
\bea
t_{\rm disr,0}\simeq& 2.7(\gamma U_{7})^{-1}\bar{\lambda}_{0.5}^{-1}\hat{\rho}^{1/2}S_{\max,9}^{1/2}a_{-5}^{2}{~\rm days}
\ena
for $a_{\rm trans}<a<a_{\rm disr,max}$ where $U_{7}=U/10^{7}$.

Figure \ref{fig:adisr} (lower panel) shows the disruption time obtained for the different tensile strengths. The disruption time decreases rapidly with the heliocentric distance.

\begin{figure}
\includegraphics[width=0.4\textwidth]{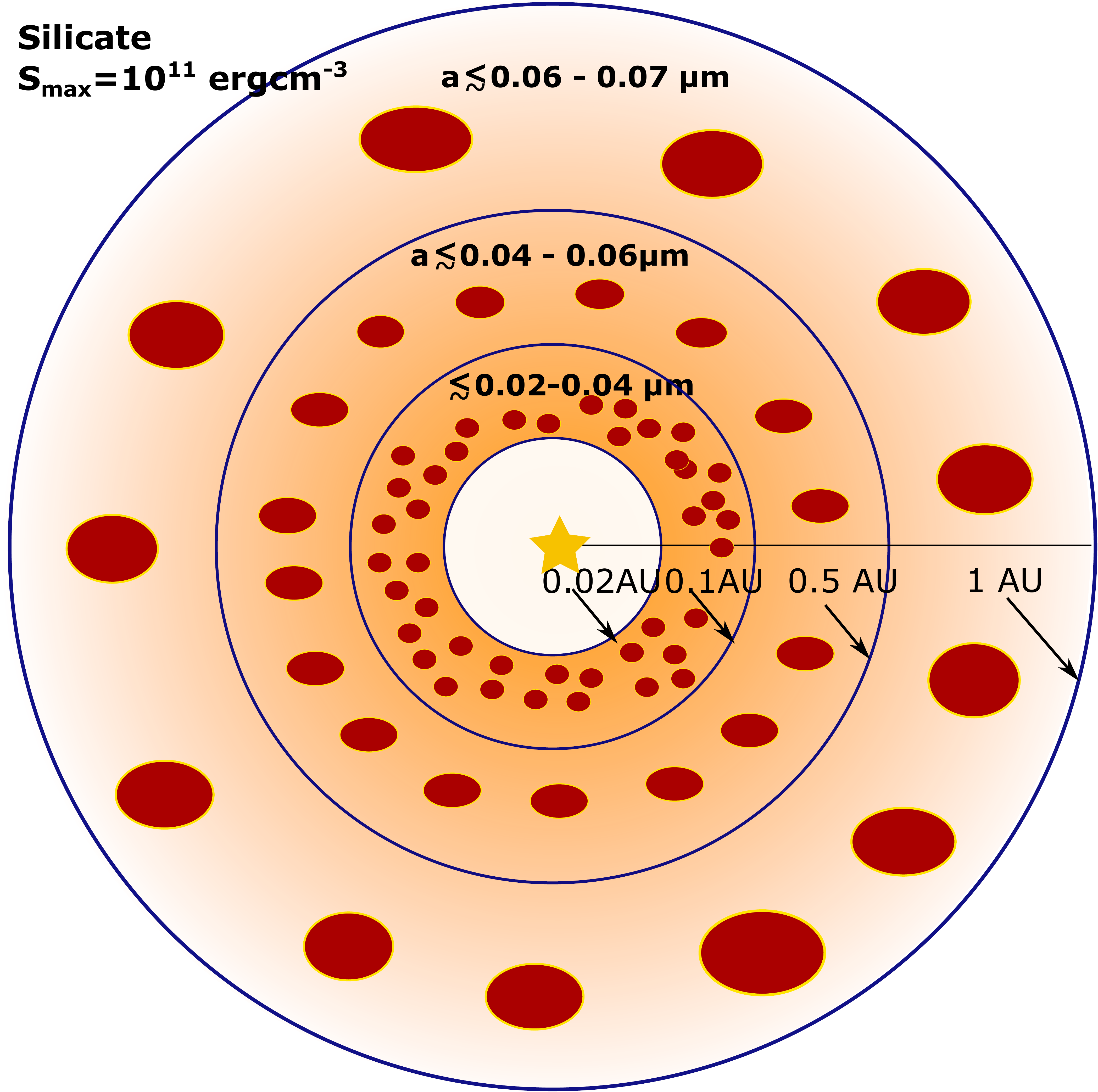}
\caption{Schematic illustration of spatial variation of grain sizes in the F-corona as a result of RATD: abundance of small grains increases toward the Sun.}
\label{fig:adisr_Fcorona}
\end{figure}

Our results reveal that grains within $0.1$ AU ($21 R_{\odot}$) are dominated by nanoparticles of size $a<0.01\mum$ (10 nm) due to RATD, as illustrated in Figure \ref{fig:adisr_Fcorona}). 

\section{Destruction of nanoparticles by the solar wind}\label{sec:nano}
In this section, we will show that smallest nanoparticles near the Sun produced by RATD will be efficiently destroyed by bombardment of energetic protons from the solar wind.

\subsection{Solar wind components}
The solar wind usually consists of very slow ($v\sim 200-300\km\s^{-1}$), slow ($v\sim 300-700 \km\s^{-1}$), and fast ($v>700 \km\s^{-1}$) solar winds (see, e.g., \cite{1981A&A....95..373M}). The very slow winds (VSLW) are observed frequently in the inner solar region of heliocentric distance $R<1\AU$ such as by Helios 1/2, but are rarely near the Earth and the density of VSLW is much higher than that of fast solar winds (\citealt{2016JGRA..121.2830S}).

The number density of solar wind protons varies with the heliocentric distance and is given by
\bea
n_{p}(R)=n_{0}\left(\frac{R}{R_{0}}\right)^{-2},\label{eq:np}
\ena
where $n_{0}=6\cm^{-3}$ at $R_{0}=1\AU$ (see e.g., \cite{Venzmer:2018gy}). At heliodistance of $R=0.1\AU$ ($\sim 21R_{\odot}$), the proton density increases to $n_{p}= 600\cm^{-3}$.


\subsection{Nonthermal sputtering}
Bombardment of energetic protons gradually erodes the grain surface via the sputtering mechanism (see e.g., \citealt{2015ApJ...806..255H}; \citealt{Hoang:2020kh}). 

To calculate the sputtering time, we assume a uniform distribution of proton velocity, and the mean velocity of solar wind (sw) protons is $\bar{v}_{\rm sw}\sim 400\km\s^{-1}$. Let $Y_{\rm sp}$ be the sputtering yield induced by the proton bombardment. The characteristic timescale, $t_{\rm nsp}$, of grain destruction by nonthermal sputtering for a spherical grain of size $a$ is defined by 
\bea
t_{\rm nsp}&=&\frac{a}{da/dt}=\frac{4\rho a}{n_{p}m_{p}\bar{v}_{\rm sw}Y_{\rm sp}\bar{A}_{\rm sp}}\nonumber\\&\simeq& 3486\left(\frac{\hat{\rho}}{n_{2}v_{2}}\right)\left(\frac{12}{\bar{A}_{\rm sp}}\right)\left(\frac{a}{10^{-7}\cm}\right) \left(\frac{0.5}{Y_{\rm sp}}\right) \rm days,~~~~\label{eq:tau_sp}
\ena
where $n_{2}=n_{\rm p}/(100\cm^{-3})$, and $\bar{A}_{sp}$ is the mean atomic mass of sputtered atoms from the grain (see e.g., \citealt{Hoang:2020kh}).

At $R\sim 0.1\AU$ ($\sim 21R_{\odot}$), $n_{p}\sim 600\cm^{-3}$ (see Eq. \ref{eq:np}), the sputtering time is $t_{\rm nsp}=1093 (a/10^{-7}\cm)$ days, assuming $Y_{\rm sp}=0.5$. Thus, the sputtering time is much longer than the rotational disruption time $t_{\rm disr}$ for grains of $a>a_{\rm dirs}$ (see the right panel of Figure \ref{fig:adisr}).

To study if nonthermal sputtering can be fast enough to destroy nanoparticles before they are dragged to the sublimation zone by the Poynting-Robertson (P-R) drag, we calculate the ratio of nonthermal sputtering time to the P-R drag time (see Eq. \ref{eq:tau_PR}):
\bea
\frac{t_{\rm nsp}}{t_{P-R}}&\approx& 0.03\left(\frac{\langle Q_{\rm pr}\rangle}{0.1}\right)\left(\frac{500\cm^{-3}}{n_{p}}\right)\left(\frac{300~\km\s^{-1}}{\bar{v}_{\rm sw}}\right)\times\nonumber\\
&&\times \left(\frac{0.5}{Y_{\rm sp}}\right)\left(\frac{\AU}{R}\right)^{2}.\label{eq:tnsp_tPR}
\ena
where $\langle Q_{\rm pr}\rangle$ is the averaged radiation pressure cross-section efficiency which is a function of the grain size.

For nanoparticles in the solar radiation field with $\langle Q_{\rm pr}\rangle\lesssim 0.1$ (see Figure \ref{fig:Qpr}), Equation (\ref{eq:tnsp_tPR}) reveals that nanoparticles are rapidly destroyed by nonthermal sputtering before dragged into the sublimation zone by the P-R drag. The efficiency of nonthermal sputtering is increased with decreasing the heliocentric distance due to the increase of proton density (see Equation \ref{eq:np}).

We note that for nanoparticles, rotational disruption by stochastic mechanical torques by the solar wind is less efficient than nonthermal sputtering because energetic protons tend to pass through the grain (\citealt{Hoang:2020kh}). \cite{1993JGR....9818951M} studied the rotational bursting of circumsolar dust caused by solar winds protons using the "Paddack effect"--"windmill" (\citealt{1975GeoRL...2..365P}). However, \cite{1993JGR....9818951M} did not take into account the effect of proton passage and assumed a rather low tensile strength of $S_{\max}=10^{6}\erg\cm^{-3}$ for circumsolar dust, which is quite low for dust grains of compact structures. The rotational disruption by regular mechanical torques due to the interaction of the solar wind with irregular grains (\citealt{2007ApJ...669L..77L}; \citealt{2018ApJ...852..129H}) is expected to be more efficient. A detailed study of this effect is presented elsewhere.

\subsection{Extended dust-free-zone and decrease of dust mass due to sputtering}

\begin{figure*}
\includegraphics[width=0.5\textwidth]{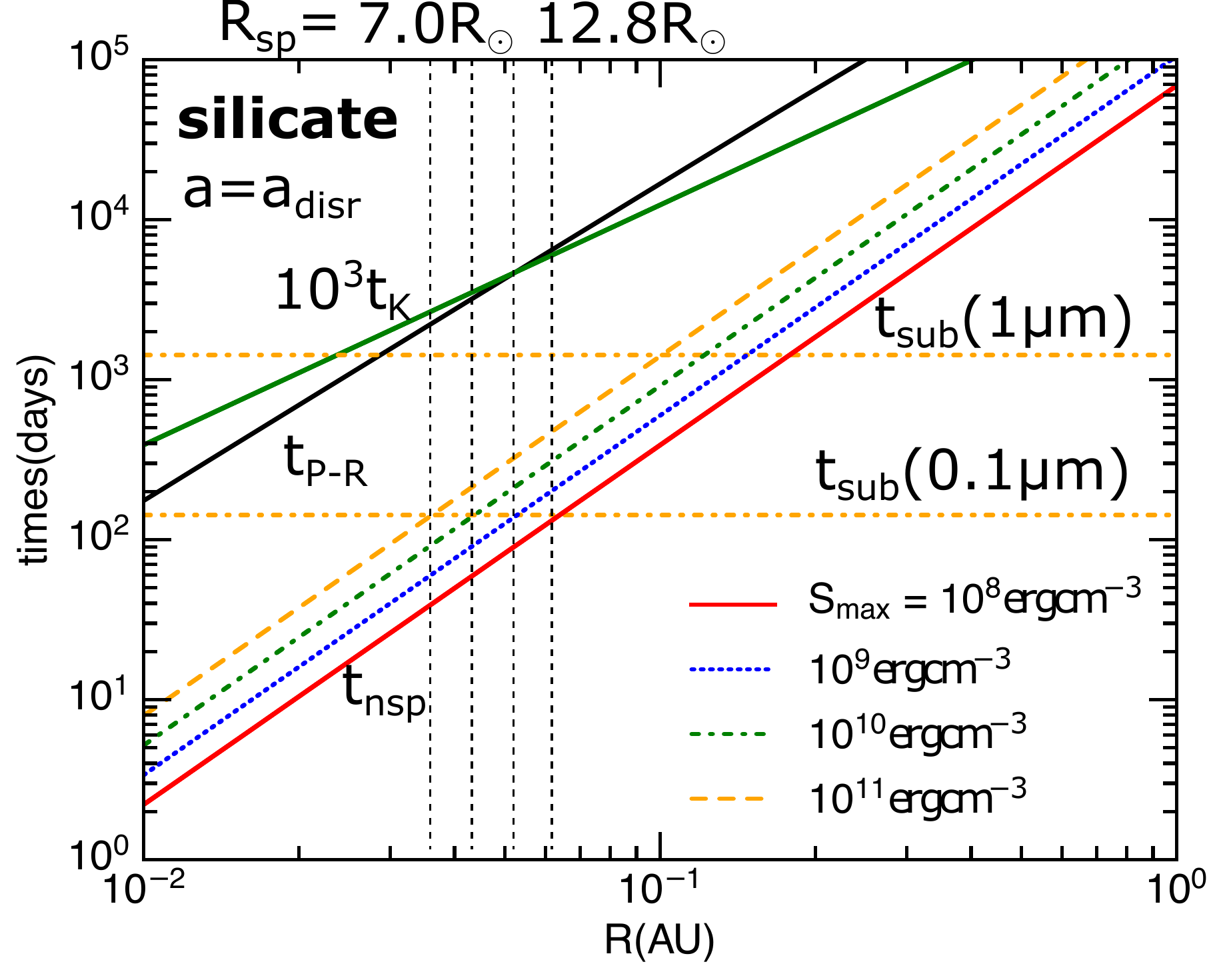}
\includegraphics[width=0.5\textwidth]{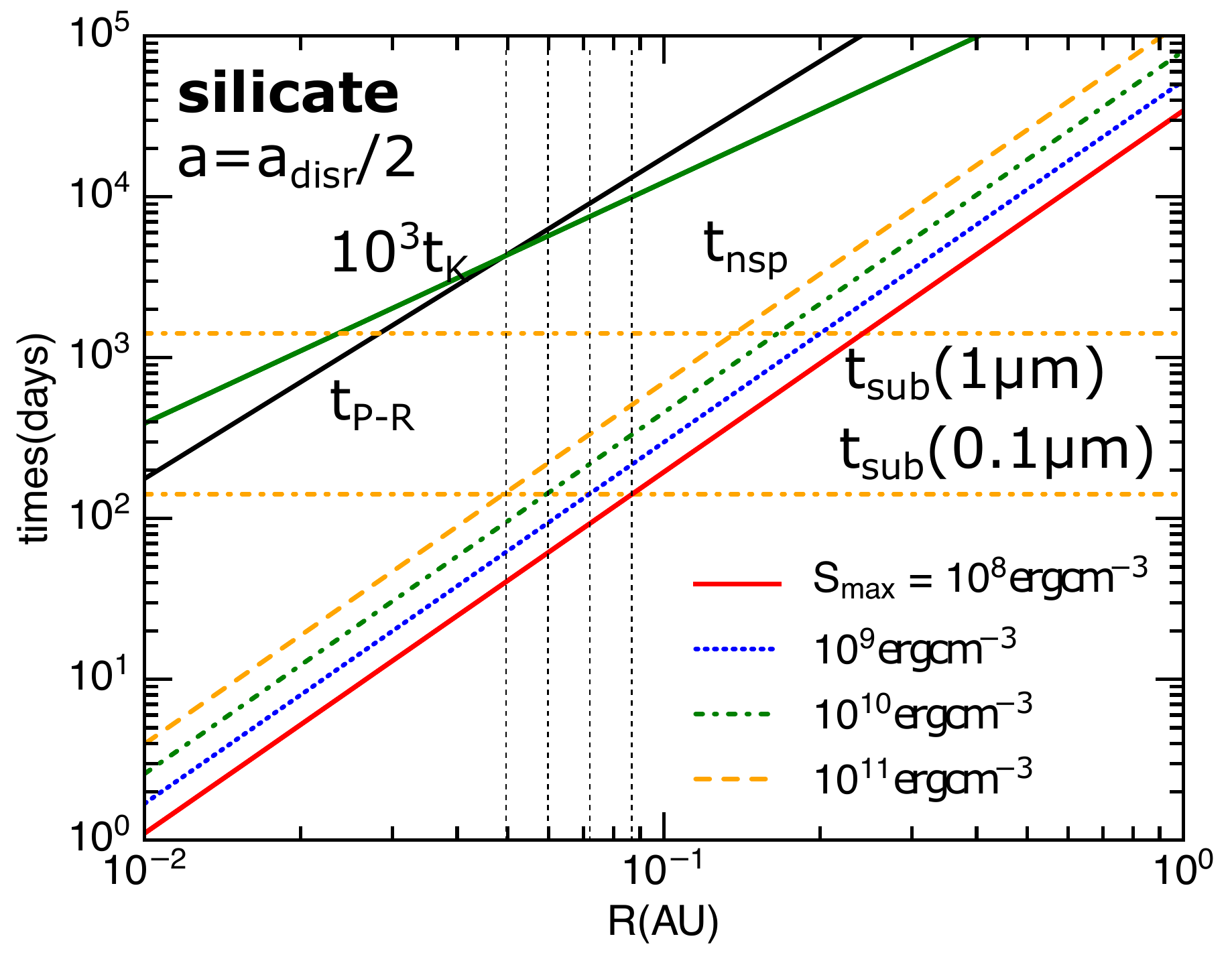}
\includegraphics[width=0.5\textwidth]{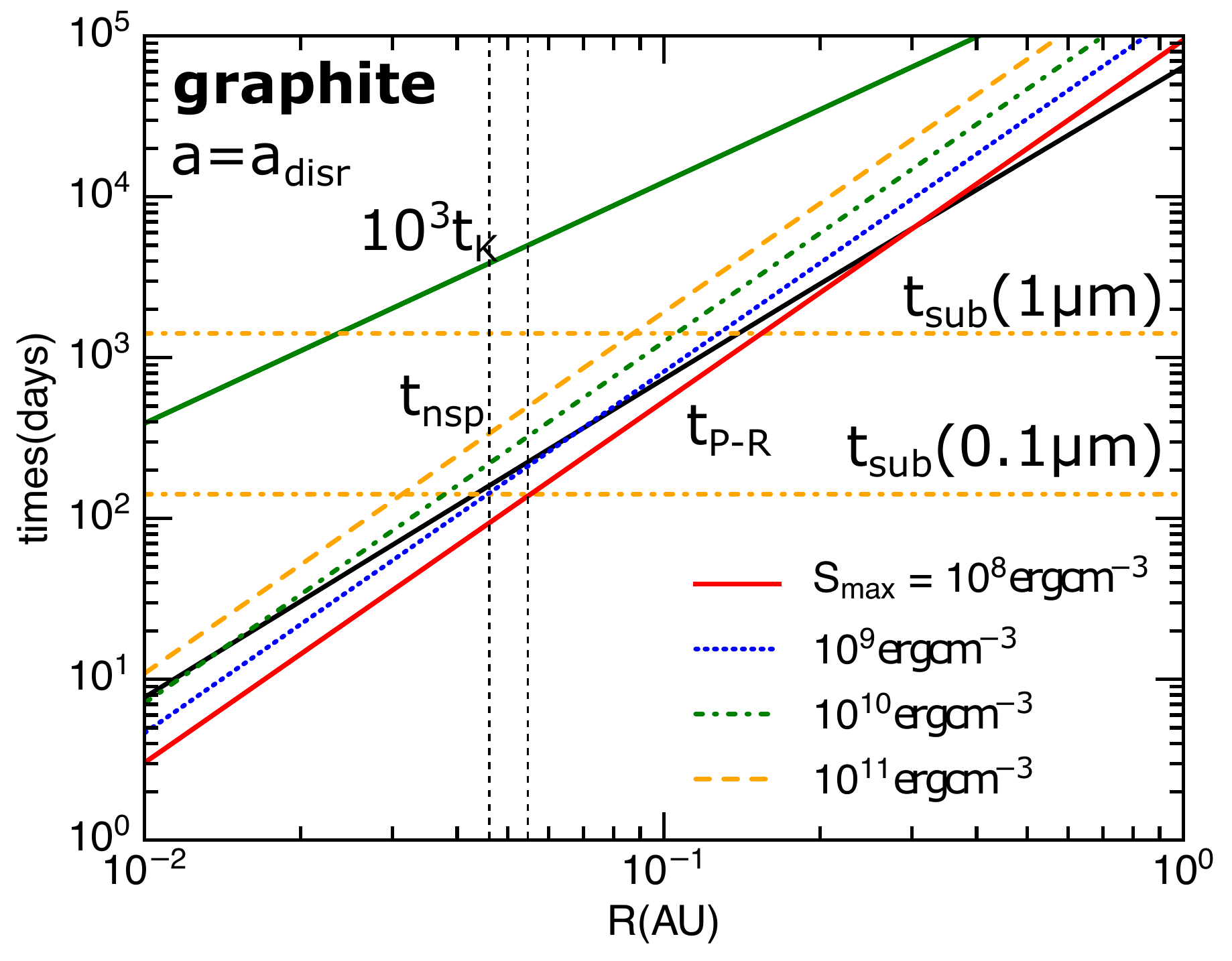}
\includegraphics[width=0.5\textwidth]{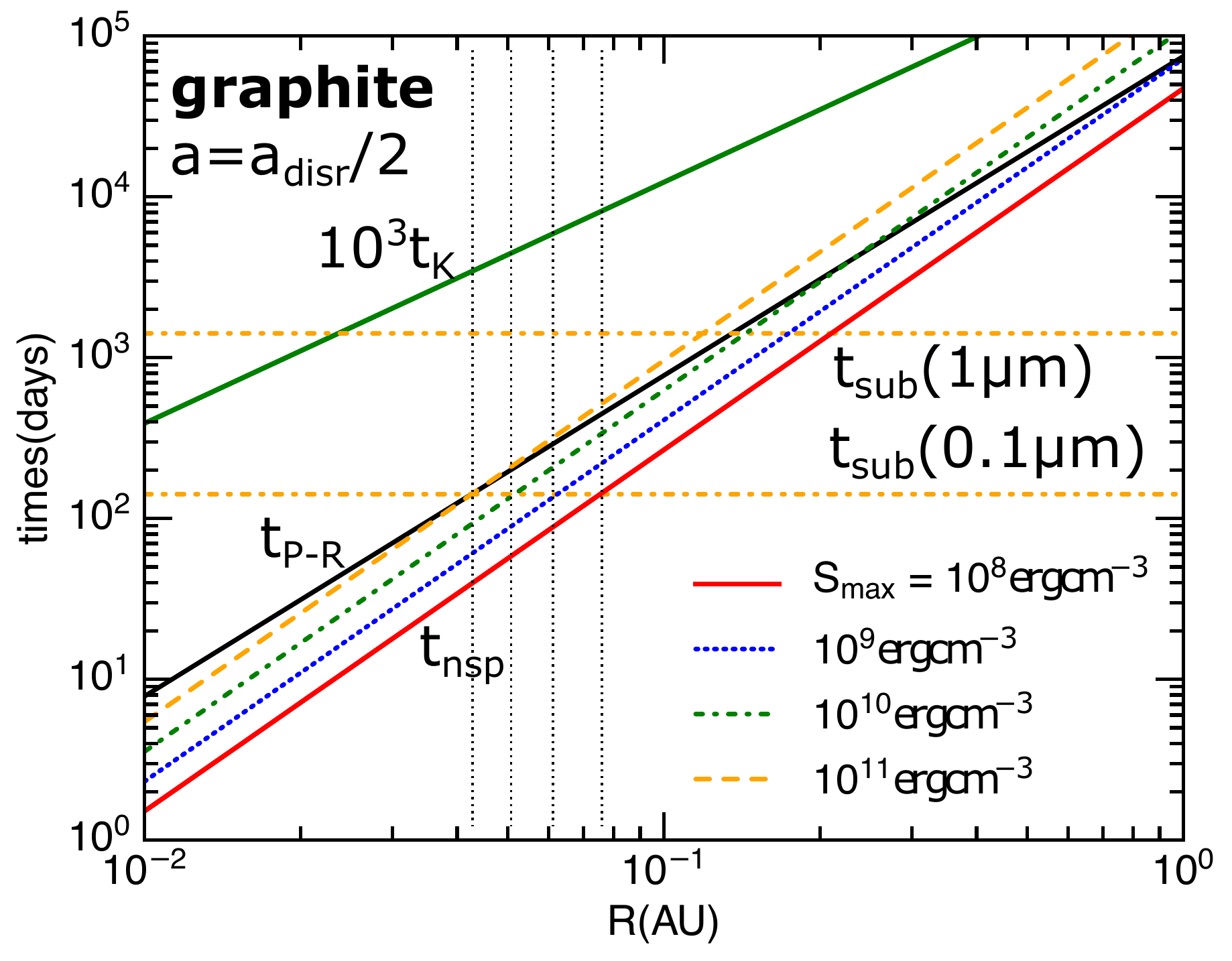}
\caption{Comparison of the characteristic timescales due to various processes, including nonthermal sputtering ($t_{\rm nsp}$), the P-R drag ($t_{P-R}$), Keplerian time $t_{K}$ for grains of sizes $a=a_{\rm disr}$ and various $S_{\rm max}$ (left panels). The sublimation time ($t_{\rm sub}$) for classical grains $a=0.1\mum$ and $1\mum$ are shown for comparison. Right panels show results for $a=a_{\rm disr}/2$. The dashed vertical lines denote the location where $t_{\rm nsp}=t_{\rm sub}(a=0.1\mum)$.}
\label{fig:timescale}
\end{figure*}

Figure \ref{fig:timescale} compares the nonthermal sputtering time evaluated at $a=a_{\rm disr}$ and $a=a_{\rm disr}/2$ to other relevant timescales. The sputtering yield for protons of $\bar{v}_{\rm sw}$ is calculated as in \cite{Hoang:2020kh}. The P-R drag time is calculated using Equation (\ref{eq:tau_PR}) where $\langle Q_{\rm pr}\rangle$ is a function of the grain size as shown in Figure \ref{fig:Qpr}. The sublimation time $t_{\rm sub}$ for two grain sizes at $T_{d}=1500\K$ is shown for comparison (see Eq. \ref{eq:tausub}). 

The dashed vertical lines denote the location of the extended dust-free-zone determined by nonthermal sputtering $R_{\rm sp}$. The sputtering radius is $R_{\rm sp}\sim 7-13R_{\odot}$ for grains of $S_{\rm max}\sim 10^{9}-10^{11}\erg\cm^{-3}$ for silicate and $8R_{\odot}$ for $S_{\rm max}\sim 10^{9}\erg\cm^{-3}$ for graphite. Note that graphite is not found in the interplanetary dust, the choice of graphite here is an example of a highly absorptive dust, such as hydrogenated amorphous carbon.

In Figure \ref{fig:timescale}, the region where $t_{\rm nsp}(a=a_{\rm disr})\lesssim t_{\rm sub}(T_{d}=1500\K)$ is considered an extended dust-free-zone. Note that the classical dust-free-zone is defined by thermal sublimation where grains are destroyed rapidly on a timescale of $t_{\rm sub}(T_{d}=1500\K)$. Thus, if nonthermal sputtering could destroy grains in the same time beyond this zone, then, the dust-free-zone is extended.

Beyond the extended dust-free-zone, nanoparticles of $a<a_{\rm disr}$ can be depleted by nonthermal sputtering because $t_{\rm nsp}(a)<t_{\rm sub}$. As a result, the mass loss of dust via nonthermal sputtering is decreasing outward, or the total dust mass decreases inward until the dust-free-zone. 



Let $a_{\rm nsp}$ be the critical size of nanoparticles that are destroyed by nonthermal sputtering. One can determine $a_{\rm nsp}$ by setting $t_{\rm nsp}(a_{\rm nsp})=t_{\rm sub}$. All grains smaller than $a_{\rm nsp}$ are destroyed by nonthermal sputtering. The value of $a_{\rm nsp}$ will increase with decreasing $R$ due to the increase of the proton density (see Eq. \ref{eq:np}).

\begin{figure}
\includegraphics[scale=0.45]{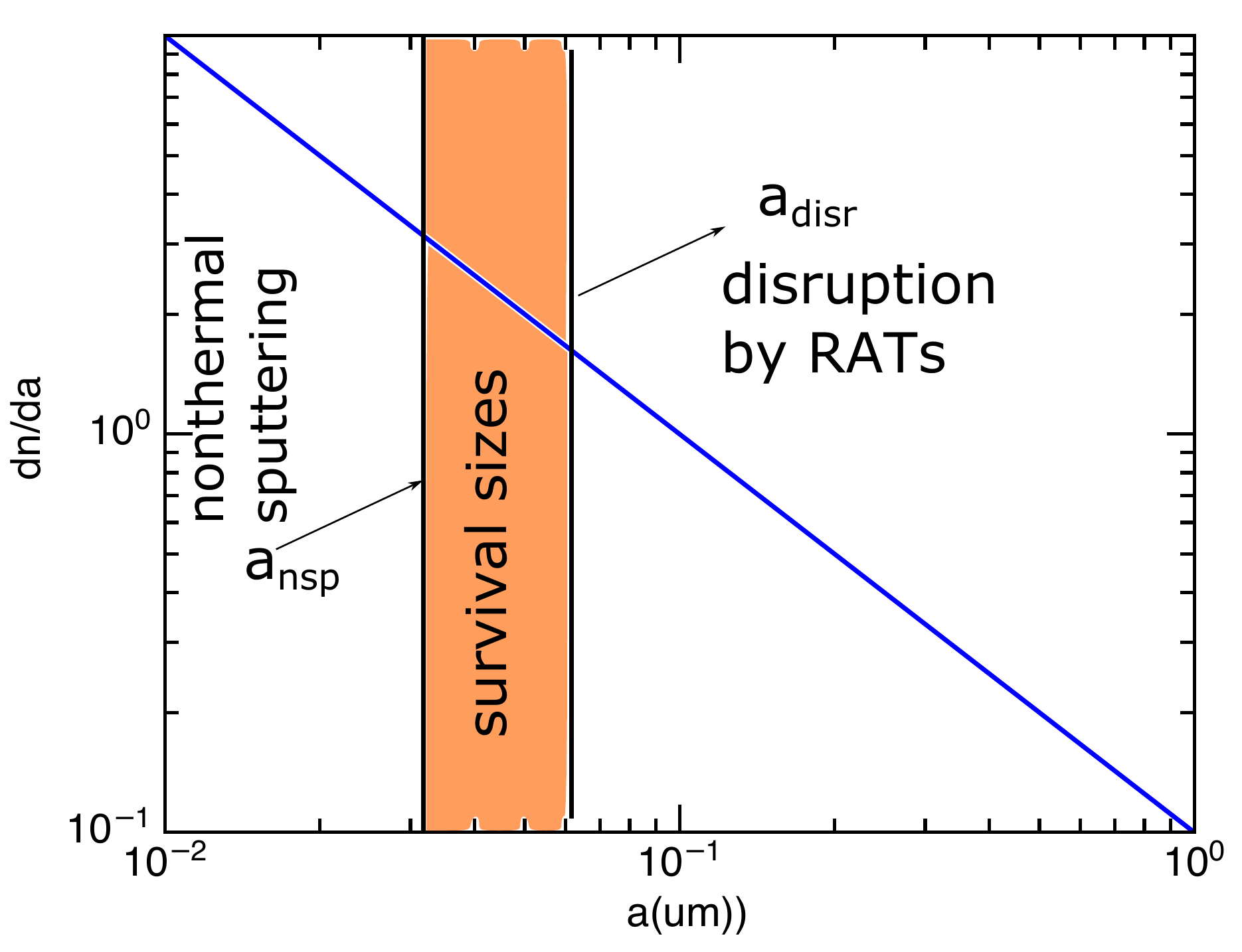}
\caption{Grain size distribution as a result of RATD and sputtering. Grains larger than $a_{\rm disr}$ are disrupted into grains of $a<a_{\rm disr}$, while nonthermal sputtering removes grains of $a<a_{\rm nsp}$.}
\label{fig:dnda}
\end{figure}

Figure \ref{fig:dnda} illustrates the grain size distribution ($dn/da\propto a^{-3.5}$) as a result of the RATD mechanism and nonthermal sputtering. Only small grains within a narrow range (shaded area) could survive in the F-corona.

Let $f_{\rm high-J}$ be the fraction of grains that can be spun-up to suprathermal rotation and then be disrupted by RATD (see more in Section \ref{sec:discuss}. Then, we can calculate the decrease of the dust mass as a function of heliocentric distance as follows:
\bea
\frac{M_{d}(R)}{M_{d}(R,0)}&=&\frac{\int_{a_{\rm nsp}}^{a_{\rm disr}} f_{\rm high-J}(4\pi a^{3}\rho /3) (dn/da)da}{M_{d}(R,0)}\nonumber\\
&&+\frac{(1-f_{\rm high-J})M_{d}(0)}{M_{d}(0)},
\ena
where $M_{d}(R,0)$ is the original dust mass in the absence of sputtering
\bea
M_{d}(R,0)=\int_{a_{\rm min}}^{a_{\rm max}} \left(\frac{4\pi a^{3}\rho }{3}\right) \left(\frac{dn}{da}\right)da
\ena
where a power-law grain size distribution of $dn/da = Ca^{-3.5}$ with $a_{\rm min}=3$~\AA~ is adopted. 

\begin{figure*}
\includegraphics[width=0.5\textwidth]{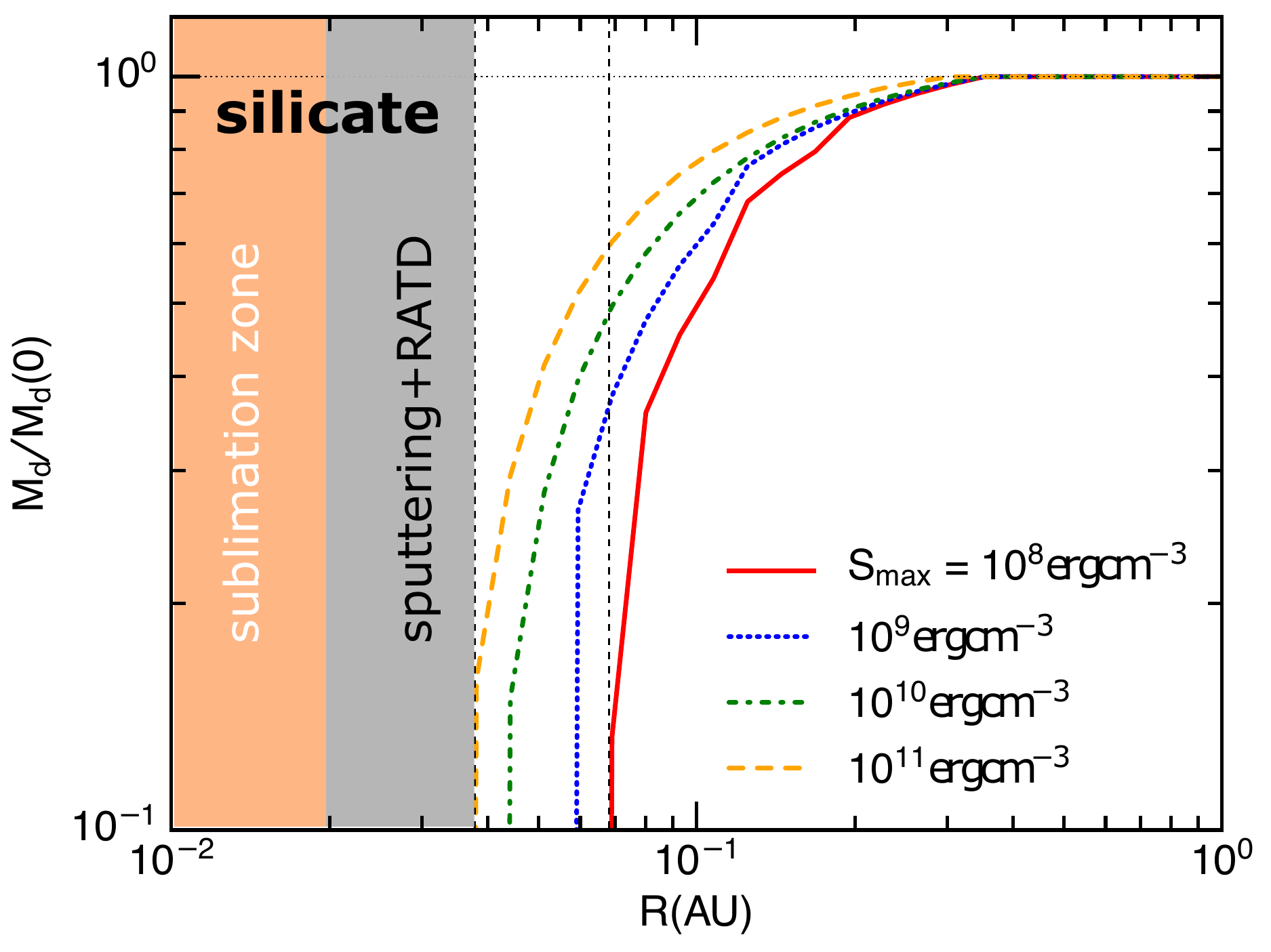}
\includegraphics[width=0.5\textwidth]{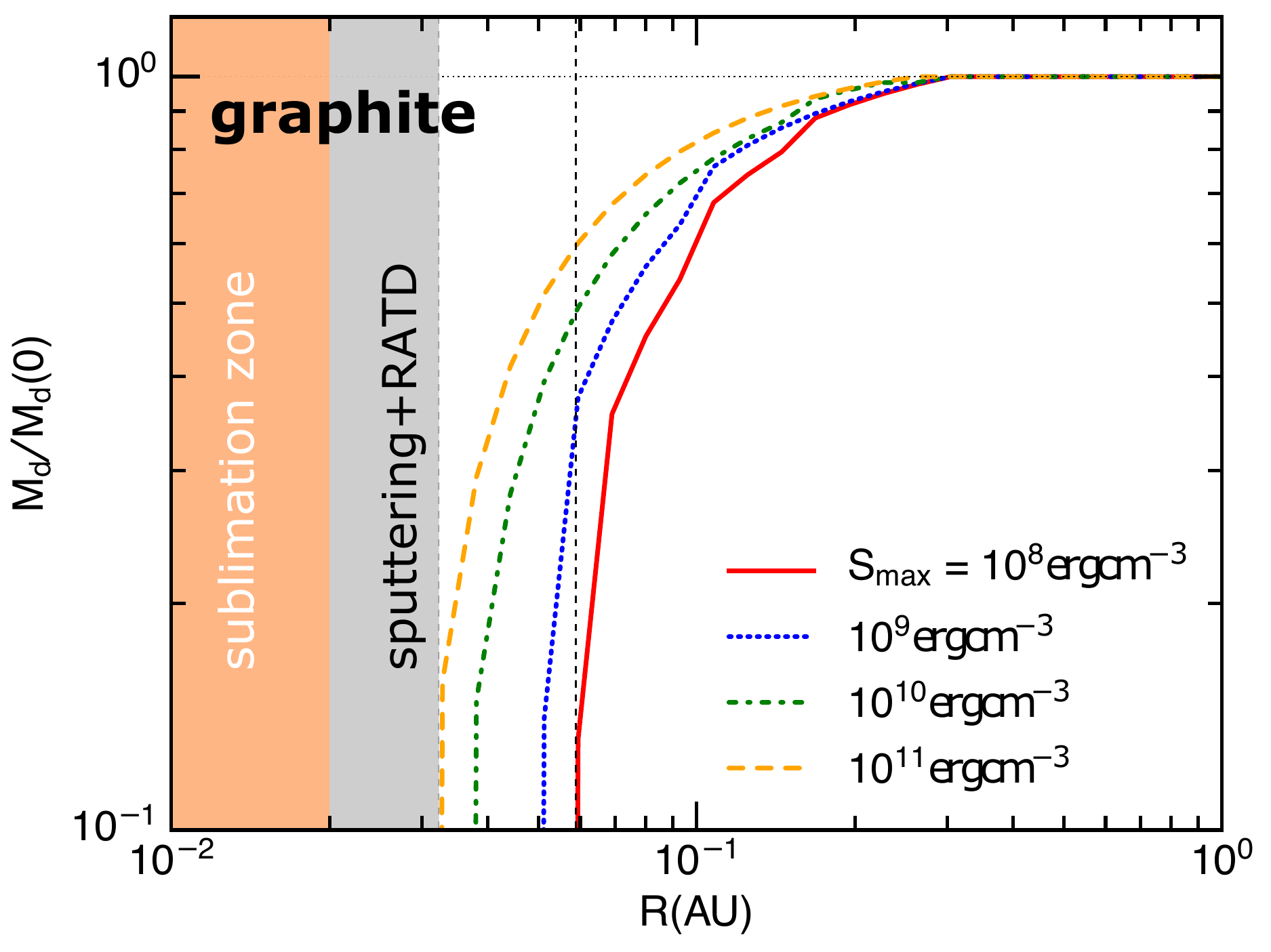}
\caption{Decrease of the dust mass ($M_{d}/M_{d}(0)$) with heliocentric distance due to RATD and nonthermal sputtering by the solar wind for silicate (left panel) and graphite (right panel) grains where $M_{d}(0)$ is the original dust mass. Four values of the tensile strength $S_{\max}=10^{8}-10^{11}\erg\cm^{-3}$ are considered. The dust mass decreases toward the Sun, starting from $\sim 0.2\AU$ (for highest strength) and reaches the extended dust-free-zone predicted by RATD and sputtering effects (gray shaded area).}
\label{fig:Md}
\end{figure*}

Figure \ref{fig:Md} shows the variation of the relative dust mass $M_{d}/M_{d}(0)$ vs. the heliocentric distance as a result of RATD and sputtering for the different tensile strength and perfect disruption with $f_{\rm high-J}=1$. We see that $M_{d}/M_{d}(0)$ starts to decrease considerably from $R\sim 0.2$ AU ($42R_{\odot}$), and a significant mass loss occurs at $R\lesssim 0.03\AU ~(6R_{\odot})$, which suggests an extended dust-free-zone. Grains made of weak material are more efficiently disrupted by RATD and experience larger mass loss.

Note that in the absence of RATD, nonthermal sputtering still can remove nanoparticles, but the fraction of depleted mass is negligible because the dust mass is mostly dominated by largest grains of $a\sim a_{\rm max}$. The effect of RATD converts large grains $a>a_{\rm disr}$ into smaller sizes and increases the abundance of nanoparticles. As a result, the amount of dust mass depleted by sputtering is significantly enhanced. However, it is uncertain whether RATD is efficient for very large grains (VLGs) of $a>\bar{\lambda}/0.1\sim 9\mum$ due to the lack of numerical calculations of RATs. If circumsolar dust is dominated by VLGs, the effect of RATD and nonthermal sputtering is less efficient. 

\section{Discussion}\label{sec:discuss}
\subsection{Effects of RATD and nonthermal sputtering}
We have shown that RATD rapidly breaks large dust grains into nanoparticles. This results in the increase in the abundance of small grains relative to large ones toward the Sun. Subsequently, nonthermal sputtering by the solar wind destroys smallest grains. Due to the increase of the solar wind density with decreasing the heliocentric distance, nonthermal sputtering rate is rapidly increased. As a result, dust grains can be completely removed by nonthermal sputtering near the Sun, which extends the dust-free-zone implied by thermal sublimation. We found that the extended dust-free-zone is located within a radius of $R_{dfz}\sim 6-12R_{\odot}$ (depending on $S_{\rm max}$, which is much larger than the classical one defined by thermal sublimation of $4-5R_{\odot}$. 

\begin{figure}
\includegraphics[width=0.5\textwidth]{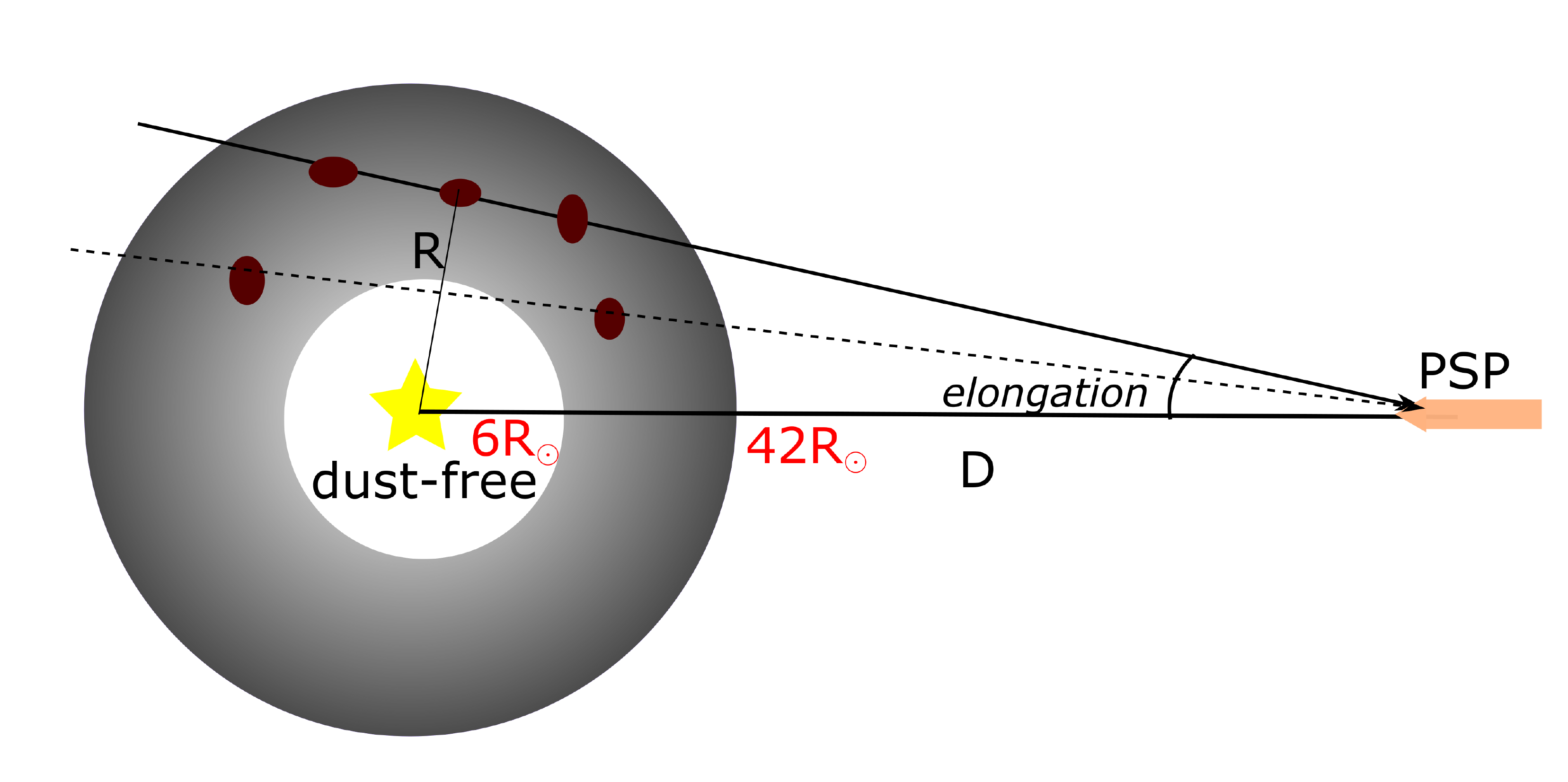}
\caption{Schematic illustration of the F-corona predicted by RATD and nonthermal sputtering which would be observed by the PSP at heliocentric distance of $D=0.3$ AU and various elongation angles. The F-corona decrease is illustrated by a radial gradient, starting from $42R_{\odot}$ to the edge of the extended dust-free-zone at $6R_{\odot}$.}
\label{fig:Fcorona}
\end{figure}

Our results also suggest that F-corona dust as well as dust in the inner solar system ($R<1\AU$) mostly contain nanoparticles of size $a\sim 1-10$ nm (see Figure \ref{fig:dnda}). This is a plausible explanation for the existence of nanodust detected by in-situ measurements (see, e.g.,  \citealt{Mann:2007em}; \citealt{Mann:2017ca}; \citealt{Ip:2019ek}). This RATD mechanism is more efficient than collisional fragmentation previously thought (see e.g., \citealt{Mann:2007em} for discussion of various mechanisms to form nanodust in the inner solar system).   

In-situ measurements by dust detector on board the Helio spacecrafts reported the F-corona decrease at heliocentric distances between $D=0.3-1$ AU (\citealt{1985ASSL..119..105G}). This is thought due to the mutual collisions that makes grains smaller and decrease of the forward scattering cross-section. However, the RATD appears to be more efficient in producing small grains due to its short timescale.

Note that RATD is valid for grains of $\bar{\lambda}/a>0.1$. For larger grains of $a>\bar{\lambda}/0.1\sim 9\mum$, RATs of such very large grains are not yet available due to the lack of numerical calculations because it requires expensive computations to achieve reliable results of RATs for grains of $2\pi~a/\lambda\gg 1$ (see e.g., \citealt{Draine:2004p6718}). Expecting the decrease of RATs with increasing $a$, the disruption may still be important when the decrease of RATs is compensated by the increase of the radiation energy density.

\subsection{Implications for the first-year results by the Parker Solar Probe}

The first-year results from the PSP at heliocentric distances of $D=0.16-0.25$ AU ($34.3- 53.7R_{\odot}$) reveal the gradual decrease the F-corona \citep{Howard:2019ih}. With the elongation $\epsilon\sim 15-20^{\circ}$, one can estimate the corresponding elongation in solar radii $R_{\epsilon}\sim (0.166-0.336)\sin(15^{\circ})\AU\sim 9-19R_{\odot}$.

To explain the first-year PSP results, let us first recall that the intensity of scattered sunlight by dust depends on the dust scattering cross-section as follows:
\bea
I_{\rm sca}&\propto& \sigma_{\rm sca}(\lambda)=\int Q_{\rm sca}(a,\lambda)\pi a^{2}\left(\frac{dn}{da}\right)da,\nonumber\\
&\propto& \int \left(\frac{Q_{\rm sca}(a,\lambda)}{a}\right) \left(\frac{\pi a^{3}dn}{da}da\right),\label{eq:Isca}
\ena
where the term in the second bracket describes the dust mass.

Equation (\ref{eq:Isca}) reveals that the decrease of dust mass by sputtering and RATD as shown in Figure \ref{fig:Md} directly reduces the observed intensity. Moreover, the decrease of grain sizes by RATD also decreases the scattering efficiency $Q_{\rm sca}$ because in the small grain limit ($a\ll \lambda$), $Q_{\rm sca}(a,\lambda)\sim (2\pi a/\lambda)^{4}$. Thus, the joint effect of RATD and sputtering would decrease the brightness of the F-corona.

Figure \ref{fig:Fcorona} illustrates the F-corona as a result of RATD and nonthermal sputtering, which would be observed with the PSP. The extended dust-free-zone is located at $R=6R_{\odot}$. Beyond this radius, the F-corona decreases with the radius, starting from $42R_{\odot}$ to the dust-free-zone.
 
Note that the sublimation radius given by Equation (\ref{eq:Rsub_sil}) reveals the range of the dust-free-zone between $4-5R_{\odot}$ for silicate grains of sizes $a\sim 0.1-0.001\mum$. Therefore, even with the effect of RATD, thermal sublimation alone cannot explain the thinning-out of circumsolar dust observed from $R_{\epsilon}\lesssim 19R_{\odot}$. However, our results shown in Figure \ref{fig:Md}) indicate that the joint effect of RATD and nonthermal sputtering could successfully explain the gradual decrease of F-corona toward the Sun. Moreover, the PSP's observation is consistent with our result with the largest value of $S_{\max}$ because this model predicts the F-corona decrease from a closest distance of $19R_{\odot}$.
 


The PSP is planned to undergo 24 orbits around the Sun. The latest orbits (22-24) will reach the closest distance of $10.74R_{\odot}$ (\citealt{2016SSRv..204....7F}). Previous studies predict that the dust-free-zone is between $4-5R_{\odot}$, which cannot be confirmed with the PSP. We found that the joint action of RATD and sputtering increase the radius of dust-free-zone to $7-11R_{\odot}$ (see Figure \ref{fig:timescale}). This would be tested with the upcoming orbits of the PSP. 

Finally, it is worth to mention that the variation of the grain size distribution by RATD affects the effect of radiation pressure. Indeed, as grains become smaller, the radiation cross-section efficiency $\langle Q_{\rm pr}\rangle$ decreases with decreasing grain size (see Figure \ref{fig:Qpr}). Thus, the ratio of radiation force to gravity $\beta=F_{\rm rad}/F_{\rm gra}$ becomes smaller than unity (see Equation \ref{eq:beta}), such that radiation pressure cannot blow these nanoparticles away.

\subsection{Effects of grain alignment with the magnetic field vs. the radiation direction}
We have assumed that RATs spin up grains to their maximum rotational angular velocity and ignored the effect of grain alignment. Thus, our obtained results  (e.g., Figure \ref{fig:Md}) are considered an upper limit of the rotational disruption effect. Previous studies show that in addition to spin up, RATs induce grain alignment along the magnetic field direction, which is usually referred to as B-RAT. Moreover, subject to the intense solar radiation, the alignment axis may change from the magnetic field (B-RAT) to the radiation direction (i.e., k-RAT; see more details in \citealt{2019ApJ...883..122L}). 
As suggested by \cite{1958ApJ...128..664P}, the magnetic field near the Sun is almost radial. Therefore, our obtained results are weakly affected by the magnetic field because the radiation and magnetic field are nearly parallel.

We have also assumed that all aligned grains have their maximum angular momentum $\omega_{\rm RAT}$ (Eqs. \ref{eq:omega_RAT} and (\ref{eq:omega_RAT0}), which corresponds to the situation that all grains are driven to high-J attractors (\citealt{2016ApJ...831..159H}; \citealt{Hoang:2008gb}). In general, the fraction of grains on high-J attractors, denoted by $f_{\rm high-J}$, depends on the grain properties (shape, size, and magnetic properties), and $0<f_{\rm high-J}\le 1$. An extensive study for numerous grain shapes and compositions by \cite{Herranen.2021} found $f_{\rm high-J}\sim 0.2-0.7$, depending on the grain shapes. The presence of iron inclusions is found to increase $f_{\rm high-J}$ to unity \citep{2016ApJ...831..159H}. It is likely that for a given size, the grains with iron inclusions would be fully disrupted first. As a consequence, we may expect to have an increase of pure “iron” tiny grains or grains with significantly increased magnetic content and pure paramagnetic grains. For the latter fraction, thermal sublimation will be only the destruction effect for the fraction of grains without high-J attractors that RATD is not effective. At the same time, our predicted dependence will be applicable to the population of high-J grains.
More details about the relation of alignment and the RATD are given in \cite{LH:2021}. Therefore, the decrease of dust mass predicted in this paper (Fig. \ref{fig:Md}) is an upper limit and may be lower by a factor $f_{\rm high-J}$. As a result, the decrease of $M_{d}$ (F-corona) would start from a smaller heliodistance of $<42R_{\odot}$.

\subsection{Existence of circumsolar dust around $4-5R_{\odot}$ and constraining the fraction of high-J attractors}
Circumsolar dust is an ideal environment to test RATD mechanism because we have both indirect and direct measurement of dust. Observations at ultraviolet, visible, and near-infrared wavelength reveal the existence of large grains between $1-10\mum$ in the circumsolar corona around $4-5R_{\odot}$. This implies that the RATD effect is not perfect (i.e., $f_{\rm high-J}<1$), as we discussed in the previous subsection. Detailed modeling of the F-corona brightness with observations would provide a constraint on the efficiency of RATD.


\section{Summary}\label{sec:sum}
We provide an additional explanation for the F-corona decrease revealed by the Parker Solar Probe based on the RATD mechanism, which does not strongly depend on the grain composition but on the grain tensile strength (internal structure). Our main results are summarized as follows:

\begin{enumerate}

\item{} We find that RATD can rapidly break large dust grains into smaller fragments, resulting in an increase in the abundance of nanoparticles with decreasing the heliocentric distance. This RATD mechanism can explain the ubiquitous existence of nanoparticles in the inner solar system.

\item{} We study destruction of circumsolar dust via nonthermal sputtering due to bombardment of energetic protons from the solar wind and find that nonthermal sputtering is efficient in destroying smallest nanoparticles. The nonthermal sputtering is faster than the P-R drag because the radiation pressure cross-section efficiency is significantly decreased for nanoparticles.

\item{} Due to the effect of RATD and nonthermal sputtering, the dust-free-zone is extended in comparison to the sublimation zone predicted by the silicate dust.

\item{} Beyond the dust-free-zone, we find that nonthermal sputtering can remove smallest nanoparticles, which results in the decrease of dust mass toward the Sun. This effect can explain the gradual decrease of the F-corona observed by Parker Solar Probe between the heliocentric distances of $D=0.166-0.336$ AU or elongations $R_{\epsilon}\sim 9-19R_{\odot}$ at $\epsilon=15^{\circ}$.

\end{enumerate}

\acknowledgements
We thank the second anonymous referee for helpful comments that improved the presentation of our paper. TH thanks Hiroshi Kamura for helpful discussion. TH acknowledges the support by the National Research Foundation of Korea (NRF) grants funded by the Korea government (MSIT) through the Basic Science Research Program (2017R1D1A1B03035359) and Mid-career Research Program (2019R1A2C1087045). AL acknowledges the support from NSF AST 1715754  grant. KSC acknowledges the support from KASI R\&D program (2020-1-850-02). PGG acknowledges the support from MOST in Taiwan through the grant MOST 105-2119-M-001-043-MY3.

\appendix
\section{Radiation pressure cross-section}

Absorption cross-section $C_{\rm abs}$ and absorption efficiency $Q_{\rm abs}=C_{\rm abs}/\pi a^{2}$. The scattering efficiency is $Q_{\rm sca}=C_{\rm sca}/(\pi a^{2})$.

The radiation pressure cross-section efficiency is given by
\bea
Q_{\rm pr}(a,\lambda)=Q_{\rm abs}+Q_{\rm sca}(1-\cos\theta),
\ena
where $\theta$ is the scattering angle. The average cross-section efficiency is obtained by integrating over the radiation spectrum (see \citealt{2015ApJ...806..255H}):
\bea
\langle Q_{\rm pr}(a)\rangle=\frac{\int Q_{\rm pr}u_{\lambda}d\lambda}{(\int u_{\lambda}d\lambda)}
\ena
denotes the averaging over the spectrum of the solar radiation field of energy density $u_{\lambda}$.

\begin{figure*}
\includegraphics[scale=0.45]{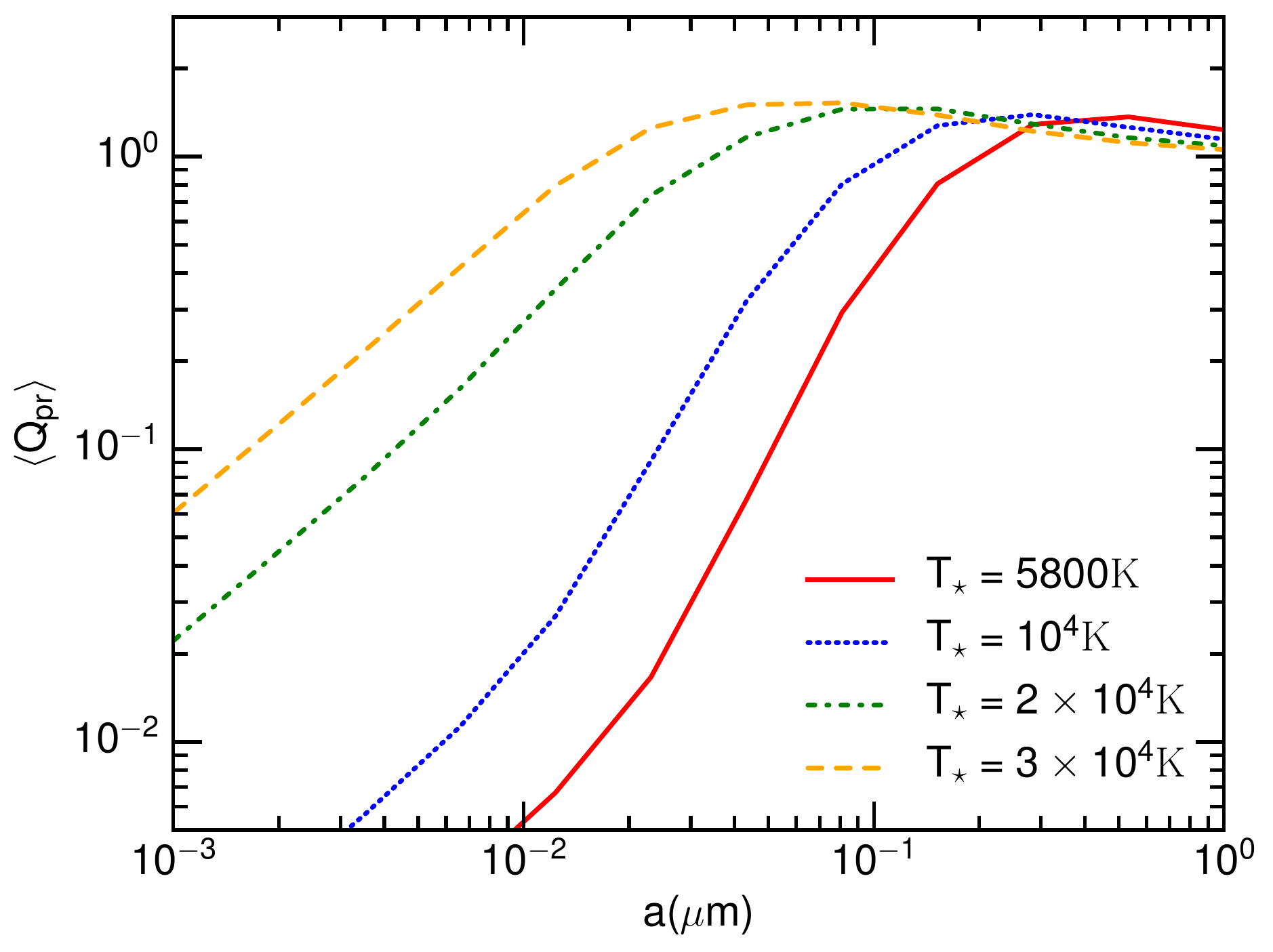}
\includegraphics[scale=0.45]{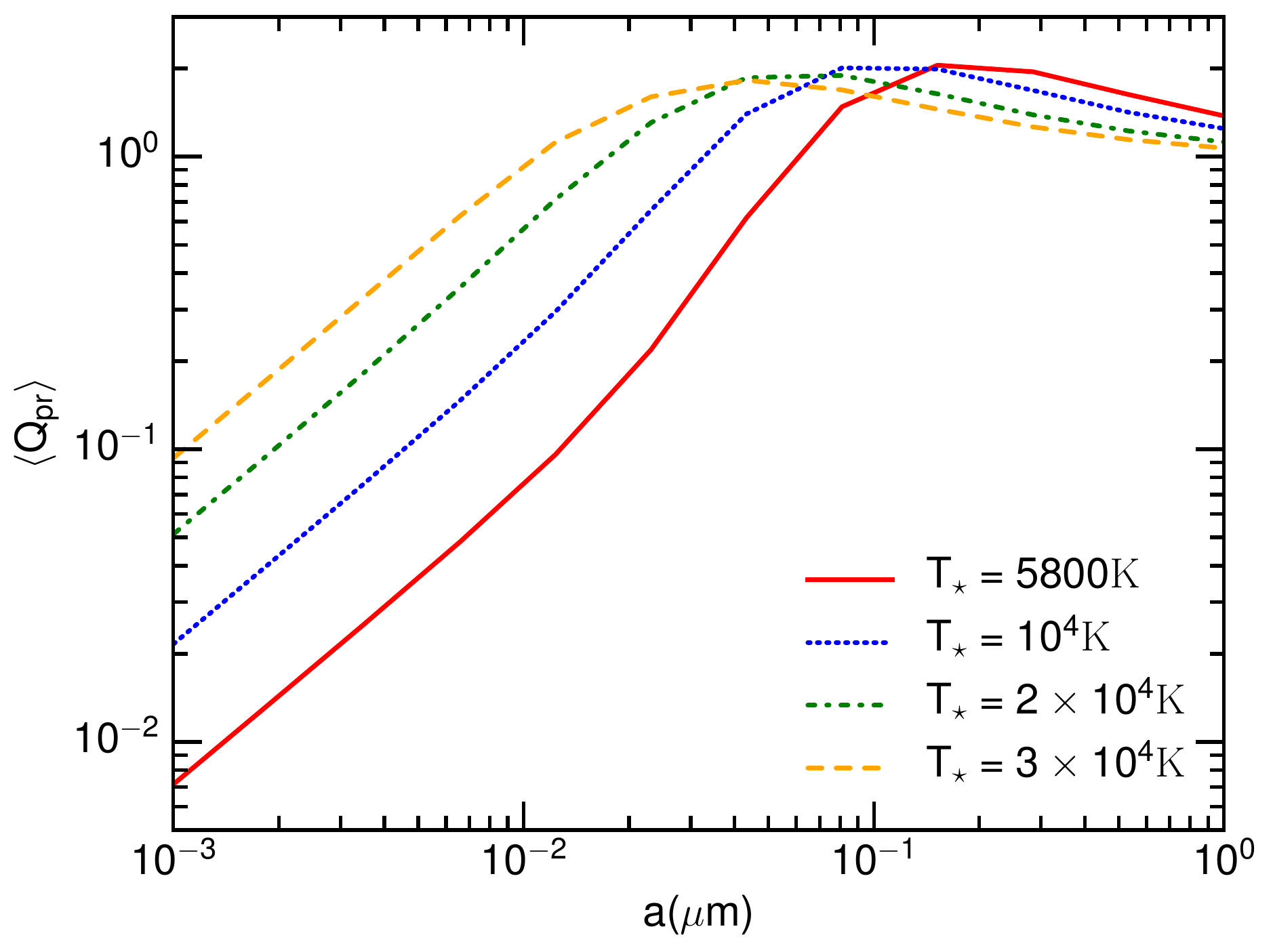}
\caption{Averaged radiation pressure cross-section efficiency as a function of the grain size assuming the stellar radiation with $T_{\star}=5800\K, 10000\K, 20000, 30000\K$ for silicate (left panel) and graphite (right panel). Rapid decrease of $\langle Q_{\rm pr}\rangle\propto a$ for $a<0.01\mum$ for the solar radiation field.}
\label{fig:Qpr}
\end{figure*}

We calculate the cross-sections for spherical grains using the Mie theory for astronomical silicate and graphite grains. Figure \ref{fig:Qpr} shows $\langle Q_{\rm pr}\rangle$ as a function of the grain size for three stellar radiation field. For the solar radiations spectrum, $Q_{\rm pr}\sim 0.1$ for $a\sim 0.01\mum$ and scales as $0.1(0.01\mum/a)$ for smaller grains.

\section{Overview of basic physics of circumsolar dust and dust-free-zone}\label{sec:overview}

\subsection{Radiation force, gravity, and Poynting-Robertson drag}
Dust grains in the interplanetary medium are subject to various processes, including solar gravity, radiative pressure, molecular force by the solar wind, and Poynting-Robertson (P-R) drag. The radiative force on a spherical grain of size $a$ located at distance $R$ from the Sun is given by
\bea
F_{\rm rad}=\frac{L_{\odot}}{4\pi R^{2}c}\langle Q_{\rm pr}(a)\rangle \pi a^{2},\label{eq:Frad}
\ena
where $L_{\odot}$ is the solar luminosity.

The ratio of radiative force to gravity force is given by
\bea
\beta=-\frac{F_{\rad}}{F_{\rm gra}}=\frac{3L_{\odot}\langle Q_{\rm pr}(a,\lambda)\rangle}{16\pi G M_{\odot}c}(\rho a)^{-1}\approx 1.91\left(\frac{\langle Q_{\rm pr}(a)\rangle}{1.0}\right)\frac{1}{\hat{\rho} a_{-5}}.\label{eq:beta}
\ena

Grains orbiting the Sun experience Poyingting-Robertson drag. \cite{Burns:1979bg} showed that the P-R force acts to gradually decrease the semimajor axis and eccentricity of the orbit of micron-sized grains around the Sun, resulting in the loss of micron-sized grains. The orbital decay time for a dust grain on circular orbit is given by
\bea
t_{\rm P-R}=\frac{R^{2}}{4\eta \langle Q_{\rm pr}\rangle}\approx 7665a_{-7} \hat{\rho}\frac{0.1}{\langle Q_{\rm pr}\rangle} \left(\frac{R}{\AU}\right)^{2} \rm days,\label{eq:tau_PR}
\ena
where $\eta=F_{0}R_{0}^{2}\pi~a^{2}/mc^{2}=2.53\times 10^{11}(\rho a)^{-1}$ with $F_{0}$ being the solar flux at distance $R_{0}$ with $m=\rho 4\pi a^{3}/3$. 

For classical grains ($a\sim 0.1-1\mum$), one has $\langle Q_{pr}\rangle\sim 1$, and the P-R drag timescale is shorter for smaller grains. For nanoparticles ($a<0.01\mum$) in the solar radiation field, the average radiation pressure decreases with the grain size as $\langle Q_{pr}\rangle\propto a$ (see Figure \ref{fig:Qpr}), thus the P-R effect become independent on $a$.

\subsection{Thermal sublimation and dust-free-zone}\label{sec:sub}
\subsubsection{Thermal sublimation}
Grains near the sun are heated to high temperature by solar radiation such that grains can sublimate rapidly. \cite{1989ApJ...345..230G} investigated the sublimation of dust grains using detailed balance and derived the sublimation rate for a grain of radius $a$:
\bea
\frac{da}{dt}=-n_{d}^{-1/3}\nu_{0}\exp\left(\frac{-B}{\kB T_d}\right),\label{eq:dasdt}
\ena
where $n_{d}\sim 10^{22}-10^{23}\cm^{-3}$ is the atomic number density of dust, $B$ is the sublimation energy per atom, $\nu_0 = 2\times 10^{15} \s^{-1}$ and $B/\kB=68100 -20000N^{-1/3}\K $ for silicate grains,  $\nu_0 = 2\times 10^{14} \s^{-1}$ and $B/\kB=81200-20000N^{-1/3} \K $ for carbonaceous grains with $N$ being the total number of atoms of the grain (\citealt{1989ApJ...345..230G}; \citealt{2000ApJ...537..796W}).

The sublimation time of a dust grain with size $a$ is defined as
\bea
t_{\rm sub}(T_d)=-\frac{a}{da/dt}
= an_{d}^{1/3}\nu_{0}^{-1}\exp\left(\frac{B}{\kB T_d}\right),\label{eq:tausub}
\ena
where $da/dt$ from Equation (\ref{eq:dasdt}) has been used.

Plugging the numerical parameters into the above equation, we obtain
\bea
t_{\rm sub}(T_d)=6.36\times 10^{3}a_{-5}\exp\left[68100\K\left(\frac{1}{T_d}-\frac{1}{1800\K}  \right)\right] \s\label{eq:tausub_sil}~~~~~
\ena
for silicate grains, and
\bea
t_{\rm sub}(T_d)=1.36a_{-5}\exp\left[81200\K\left(\frac{1}{T_d}-\frac{1}{3000\K}  \right)\right]\s\label{eq:tausub_gra}~~~~~
\ena
for graphite grains.

\subsubsection{Dust-free-zone}
When thermal sublimation is faster than the P-R drag time, grains are rapidly sublimated before they can be dragged inward by P-R, producing a dust-free-zone.

The equilibrium temperature of grains irradiated by solar radiation can be approximately given by
\bea
T_{d}=16.4a_{-5}^{-1/15} {U}^{1/6}\K\simeq 1916a_{-5}^{-1/15}\left(\frac{R}{R_{\odot}}\right)^{-1/3},\label{eq:Td_sil}\\
T_{d}=22.3a_{-5}^{-1/15} {U}^{1/6}\K\simeq 2376a_{-5}^{-1/15}\left(\frac{R}{R_{\odot}}\right)^{-1/3},\label{eq:Td_gra}
\ena
for silicate and graphite, respectively. These approximate formulae are valid for $U\le 10^{4}$ and $0.01<a<1\mum$ (silicates) and $0.005<a<0.15\mum$ (graphite). So, to calculate $T_{d}$ for a general $U$, we directly compute by solving Eq. as in \cite{2015ApJ...806..255H}.

One can estimate the sublimation distance by setting $T_{d}\equiv T_{\rm sub}$. Thus, the sublimation distance is then a function of the grain size
\bea
R_{\rm sub}=2\left(\frac{T_{\rm sub}}{1500\K}\right)^{3}\left(\frac{a}{0.1\mum}\right)^{-1/5}R_{\odot}.\label{eq:Rsub_sil}
\ena

Equation \ref{eq:Rsub_sil} reveals that the radius of dust-free-zone $R_{\rm sub}$ is between $3-4R_{\odot}$ for $a\sim 0.01-0.1\mum$.

The criteria for efficient sublimation is defined with respect to the period of Kepler motion of the dust grain required for the grain of initial distance $r$ to fall to $R_{f}=R_{\odot}$ (\citealt{1979PASJ...31..585M}):
\bea
t_{K}= \frac{(R/R_{\odot})^{3/2}}{[G(1-\beta)]^{1/2}}
\simeq 0.11 \left(\frac{R}{R_{\odot}}\right)^{3/2} \frac{1}{[(1-\beta)]^{1/2}}\rm days,
\ena
where $\beta=F_{\rm rad}/F_{\rm gra}$ is the ratio of radiative force to gravity force. 

\cite{1979PASJ...31..585M} adopted a ratio of $t_{\rm sub}/t_{K}\sim 10^{2}-10^{3}$ to define the dust-free-zone which is determined by this distance corresponding to $t_{\rm sub}/t_{K}\sim 10^{2}-10^{3}$ or $t_{\rm sub}\sim 10-100$ days. Due to RP drag, dust grains are dragged inward and concentrates around the dust-free-zone boundary, producing the circumsolar ring. The region with $t_{\rm sub}/t_{K}<10^{2}-10^{3}$ or $t_{\rm sub}\sim 10-100$ days is essentially dust-free-zone.

\section{Review of Radiative Torque Disruption (RATD) Mechanism}\label{sec:RATD}
\subsection{Rotation rate of irregular grains spun-up by radiative torques}
In this section, we briefly describe radiative torques (RATs) and spin-up by RATs for irregular grains for reference. A detailed description can be found in our previous works (\citealt{Hoang:2019da}; \citealt{2019ApJ...876...13H}).

Radiative torque (RAT) arising from the interaction of an anisotropic radiation field with an irregular grain is defined as
\bea
{\Gamma}_{\lambda}=\pi a^{2}
\gamma u_{\lambda} \left(\frac{\lambda}{2\pi}\right){Q}_{\Gamma},\label{eq:GammaRAT}
\ena
where $\gamma$ is the anisotropy degree of the radiation field, ${Q}_{\Gamma}$ is the RAT efficiency, and $a$ is the effective size of the grain which is defined as the radius of the sphere with the same volume as the irregular grain (\citealt{1996ApJ...470..551D}; \citealt{2007MNRAS.378..910L}).

The magnitude of RAT efficiency, $Q_{\Gamma}$, can be approximated by a power-law (\citealt{Hoang:2008gb}):
\bea
Q_{\Gamma}\sim 0.4\left(\frac{{\lambda}}{1.8a}\right)^{\eta},\label{eq:QAMO}
\ena
where $\eta=0$ for $\lambda \lesssim 1.8a$  and $\eta=-3$ for $\lambda > 1.8a$. 

Numerical calculations of RATs for several shapes of different optical constants in \cite{2007MNRAS.378..910L} find the slight difference in RATs among the realization. An extensive study for a large number of irregular shapes by \cite{2019ApJ...878...96H} shows little difference in RATs for silicate, carbonaceous, and iron compositions. Moreover, the analytical formula (Equation \ref{eq:QAMO}) is also in a good agreement with their numerical calculations. Therefore, one can use Equation (\ref{eq:QAMO}) for the different grain compositions and grain shapes, and the difference is an order of unity. We note that the above scaling (Eq. \ref{eq:QAMO}) is valid for grains of sizes $a<\lambda/0.1$ only (\citealt{2007MNRAS.378..910L}; \citealt{2019ApJ...878...96H}). Calculations of RATs for larger grains of $a>\lambda/0.1$ are not yet available due to the expensive computing time. However, 

The average radiative torque efficiency over the spectrum is defined as
\bea
\overline{Q}_{\Gamma} = \frac{\int \lambda Q_{\Gamma}u_{\lambda} d\lambda}{\int \lambda u_{\lambda} d\lambda}.
\ena

For interstellar grains with $a\lesssim a_{\rm trans}=\overline{\lambda}/1.8$, $\overline{Q}_{\Gamma}$ can be approximated to (\citealt{2014MNRAS.438..680H})
\bea
\overline{Q}_{\Gamma}\simeq 2\left(\frac{\overline{\lambda}}{a}\right)^{-2.7}\simeq 2.6\times 10^{-2}\left(\frac{\overline{\lambda}}{0.5\mum}\right)^{-2.7}a_{-5}^{2.7},
\ena 
and $\overline{Q_{\Gamma}}\sim 0.4$ for $a> a_{\rm trans}$.

Therefore, the averaged radiative torque can be given by
\bea
\Gamma_{\rm RAT}&=&\pi a^{2}
\gamma u_{\rad} \left(\frac{\overline{\lambda}}{2\pi}\right)\overline{Q}_{\Gamma}\nonumber\\
&\simeq & 5.8\times 10^{-29}a_{-5}^{4.7}\gamma_{\rm rad}U\overline{\lambda}_{0.5}^{-1.7}\erg,~~~
\ena
for $a\lesssim a_{\rm trans}$, and
\bea
\Gamma_{\rm RAT}\simeq & 8.6\times 10^{-28}a_{-5}^{2}\gamma U\overline{\lambda}_{0.5}\erg,~~~
\ena
for $a> a_{\rm trans}$, where $\overline{\lambda}_{0.5}=\overline{\lambda}/0.5\mum$

The well-known damping process for a rotating grain is sticking collisions with gas atoms, followed by thermal evaporation. Thus, for a gas with He of $10\%$ abundance, the characteristic damping time is
\bea
\tau_{\gas}&=&\frac{3}{4\sqrt{\pi}}\frac{I}{1.2n_{\rm H}m_{\rm H}
v_{\rm th}a^{4}}\nonumber\\
&\simeq& 8.74\times 10^{4}a_{-5}\hat{\rho}\left(\frac{30\cm^{-3}}{n_{\H}}\right)\left(\frac{100\K}{T_{\gas}}\right)^{1/2}~{\rm yr},~~
\ena
where $v_{\rm th}=\left(2k_{\B}T_{\rm gas}/m_{\rm H}\right)^{1/2}$ is the thermal velocity of a gas atom of mass $m_{\rm H}$ in a plasma with temperature $T_{\gas}$ and density $n_{\H}$, the spherical grains are assumed (\citealt{2009ApJ...695.1457H}; \citealt{1996ApJ...470..551D}). This time is equal to the time required for the grain to collide with an amount of gas of the grain mass.

IR photons emitted by the grain carry away part of the grain's angular momentum, resulting in the damping of the grain rotation. For strong radiation fields or not very small sizes, grains can achieve equilibrium temperature, such that the IR damping coefficient (see \citealt{1998ApJ...508..157D}) can be calculated as
\bea
F_{\rm IR}\simeq \left(\frac{0.4U^{2/3}}{a_{-5}}\right)
\left(\frac{30 \cm^{-3}}{n_{\H}}\right)\left(\frac{100 \K}{T_{\gas}}\right)^{1/2}.\label{eq:FIR}
\ena 
 
Other rotational damping processes include plasma drag, ion collisions, and electric dipole emission. These processes are mostly important for PAHs and very small grains (\citealt{1998ApJ...508..157D}; \citealt{Hoang:2010jy}; \citealt{2011ApJ...741...87H}). Thus, the total rotational damping rate for large grains by gas collisions and IR emission can be written as
\bea
\tau_{\rm damp}=\frac{\tau_{\gas}}{1+ F_{\rm IR}}.\label{eq:taudamp}
\ena

Due to intense solar radiation field of $U\gg 10^{10}$, the rotational damping by IR emission dominates over gas damping, i.e., $F_{\rm IR}\gg 1$. Therefore,
\bea
\tau_{\rm damp}\simeq 4.7a_{-5}^{2}\left(\frac{U}{10^{7}}\right)^{-2/3} \yr,\label{eq:taudamp2}
\ena
which does not depend on the gas properties. At $r=0.1\ AU$, $U\sim 5\times 10^{9}$, $\tau_{\rm damp}\sim 27 $days for $a\sim 0.1\mum$ grains.

For the solar radiation source with stable luminosity, radiative torques $\Gamma_{\rm RAT}$ is constant, and the grain velocity is steadily increased over time. The equilibrium rotation can be achieved at (see \citealt{2007MNRAS.378..910L}; \citealt{2009ApJ...695.1457H}; \citealt{2014MNRAS.438..680H}):
\bea
\omega_{\rm RAT}=\frac{\Gamma_{\rm RAT}\tau_{\rm damp}}{I},~~~~~\label{eq:omega_RAT0}
\ena
where $I=8\pi \rho a^{5}/15$ is the grain inertia moment.



\bibliography{ms.bbl}
\end{document}